\documentstyle[editedvolume,numreferences,epsf]{crckapb} 
\input psfig

\newcommand{\be}{\begin{equation}}
\newcommand{\ee}{\end{equation}}

\def\si{\sigma}

\def\vev#1{\left\langle#1\right\rangle}

\def\bXd{\dot{\bX}}
\def\bXp{\mathaccent 19 {\bX}}
\def\Xd{\dot X}
\def\Xp{\mathaccent 19 X}
\def\dprime{\mathaccent"707D}

\newcommand{\bX}{{\mathbf{X}}}
\newcommand{\ben}{\begin{equation}}
\newcommand{\een}{\end{equation}}
\newcommand{\bea}{\begin{eqnarray}}
\newcommand{\eea}{\end{eqnarray}}

\begin{opening}
\title{COSMIC DEFECTS AND COSMOLOGY}
 
\author{JO\~AO MAGUEIJO}
\institute{Theoretical Physics, The Blackett Laboratory, Imperial College\\
           Prince Consort Road, London SW7 2BZ, UK}

\author{ROBERT H. BRANDENBERGER}
\institute{Physics Department, Brown University\\
           Providence, RI, 02912, USA}

\end{opening}

\runningtitle{Cosmic Defects}

\begin{document}

\begin{abstract}
We provide a pedagogical overview of defect models of structure
formation. We first introduce the concept of topological defect,
and describe how to classify them. We then show how defects might
be produced in phase transitions in the Early Universe and approach
non-pathological scaling solutions. A very heuristic account of structure
formation with defects is then provided, following which we introduce
the tool box required for high precision calculations of CMB and LSS
power spectra in these theories. The decomposition into scalar
vector and tensor modes is reviewed, and then we introduce
the concept of unequal-time correlator. We use isotropy and causality
to constrain the form of these correlators. We finally show how these
correlators may be decomposed into eigenmodes, thereby reducing a defect
problem to a series of ``inflation'' problems. We conclude with a short
description of results in these theories and how they fare against
observations. We finally describe yet another application of topological
defects in cosmology: baryogenesis. 
\end{abstract}

\section{Introduction}

Phase transitions are ubiquitous in nature. Typically, as the temperature of a system drops below the critical temperature, the system makes a transition from a state with greater symmetry to one with less symmetry. In general, the state with less symmetry is not unique, but can lie anywhere in a so-called {\it vacuum manifold}. Depending on the topology of this vacuum manifold, defects will form during the phase transition. If the topology of the vacuum manifold admits defects, then these defects will inevitably arise during the phase transition unless the dynamics is completely adiabatic.

In the early Universe the temperature was decreasing very rapidly. On length scales larger than the Hubble radius, causality prevents the system from maintaining adiabaticity through interactions, and therefore on these scales defects will arise in any cosmological phase transition in which they are topologically allowed. The Universe has undergone several phase transitions. We are quite confident about those which occurred at lower temperatures: the confinement transition at a temperature $T \sim 10^2 {\rm GeV}$ and the electroweak symmetry breaking transition at $T \sim 10^3 {\rm GeV}$. Unified field theories of fundamental interactions predict the existence of other transitions at higher temperatures, e.g. a phase transition at $T \sim 10^{16} {\rm GeV}$ in Grand Unified Models, the supersymmetry breaking phase transition in supersymmetric models, and various compactification transitions in string (and M-) theory.

Since topological defects carry energy density, they will curve space-time and can thus act as the seeds for gravitational accretion (see e.g. Refs. \cite{TK80,Vil85,ShellVil,HK95,RB94} for comprehensive reviews). Since inside of topological defects the symmetry characteristic of the high temperature phase is unbroken, topological defects can interact in various interesting ways with the surrounding matter and can have an effect on cosmological issues such as baryogenesis (see e.g. \cite{BDH,BD,BDPT}), magnetic field generation (see e.g. \cite{DD,BZ99}), and ultra-high-energy cosmic ray production \cite{MB,PB,Sigl,BV}.

In these lectures, we first review the classification of topological defects and explain why in models with the appropriate topology, defects will inevitably form during the symmetry breaking phase transition. In Section 3 we discuss some initial applications of topological defects to cosmology. We review the domain wall and monopole problems and explain why the cosmic string model yields a promising mechanism for structure formation.

In the following sections we provide a more technical description
of how high accuracy calculations of structure formation in defect
theories are performed.  We first 
describe the details of the scalar, vector and tensor decomposition
(Section \ref{svt}). This is an invaluable tool in linear perturbation
theory. Then in Section \ref{corrs} we introduce the concept of UETC
and show their general form, assuming isotropy and scaling, but not
energy conservation. In Section~\ref{caus} we show how
causality limits further the form of the correlators, in the
large wavelength limit. These results will be important when checking
upon the numerics.  Then in Section~\ref{result}
we present the UETCs measured for cosmic strings, and highlight 
some of their features. Their remarkable novelty is the
dominance of the energy density over any other components of the
stress energy tensor. This property sets strings apart, resulting in
a dominance of scalar modes over vector and tensor modes. 
We present some conclusions on string scenarios of structure formation,
and also hybrid scenarios combining strings and inflation.

In the final section we illustrate one application of topological defects to cosmology which involves microscopic physics rather than gravitational accretion: we discuss the basics of defect-mediated baryogenesis.

\section{Defect Classification and Formation}

According to our current particle physics theories, matter at high energies and
temperatures must be described in terms of fields.  Gauge symmetries have
proved to be extremely useful in describing the standard model of particle
physics, according to which at high energies the laws of nature are invariant
under some non-abelian group of internal symmetry transformations $G = {\rm SU} (3)_c \times {\rm SU} (2)_L \times U(1)_Y$, where the first factor is the symmetry group of the strong interactions, and the second and third factors form the gauge group of the Glashow-Weinberg-Salam theory of electroweak interactions which is spontaneously broken to the $U(1)$ of electromagnetism at a scale of $T \sim 10^3 {\rm GeV}$.
 
Spontaneous symmetry breaking is induced by an order parameter $\varphi$ taking
on a nontrivial expectation value $< \varphi >$ below a certain temperature
$T_c$.  In some particle physics models, $\varphi$ is a fundamental scalar
field in a nontrivial representation of the gauge group $G$. However, $\varphi$ could also be a fermion condensate, as in the BCS theory of superconductivity.

The transition taking place at $T = T_c$ is a phase transition and $T_c$ is
called the critical temperature.  From condensed matter physics it is well
known that in many cases topological defects form during phase transitions,
particularly if the transition rate is fast on a scale compared to the system
size.  When cooling a metal, defects in the crystal configuration will be
frozen in; during a temperature quench of $^4$He, thin vortex tubes of the
normal phase are trapped in the superfluid; and analogously in a temperature
quench of a superconductor, flux lines are trapped in a surrounding sea of the
superconducting Meissner phase. 

In cosmology, the rate at which the phase transition proceeds is given by the
expansion rate of the Universe.  Hence, topological defects will inevitably be
produced in a cosmological phase transition \cite{Kibble}, provided the underlying particle physics model allows such defects.

Topological defects can be point-like (monopoles), string-like (cosmic
strings) \cite{ZelVil} or planar (domain walls), depending on the particle physics model.   Also of importance are textures \cite{Davis,Turok}, point
defects in space-time.
Topological defects represent regions in space with trapped energy density.
These regions of surplus energy can act as seeds for structure  formation.   
 
Consider a single component real scalar field with a typical symmetry breaking
potential
\be \label{pot}
V (\varphi) = {1\over 4} \lambda (\varphi^2 - \eta^2)^2  
\ee
Unless $\lambda \ll 1$ there
will be no inflation.  The phase transition will take place on a short time
scale $\tau < H^{-1}$, and will lead to correlation regions of radius $\xi <
t$ inside of which $\varphi$ is approximately constant, but outside of which
$\varphi$ ranges randomly over the vacuum manifold ${\cal M}$, the set of
values
of $\varphi$ which minimizes $V(\varphi)$ -- in our example $\varphi
= \pm \eta$.  The correlation regions are separated by domain walls, regions in
space where $\varphi$ leaves the vacuum manifold ${\cal M}$ and where,
therefore, potential energy is localized.  Via the usual gravitational
force, this energy density can act as a seed for structure.

As mentioned above, topological defects are familiar from solid state and condensed matter systems.   
The analogies between defects in particle physics and condensed matter
physics are quite deep.  Defects form for the same reason: the vacuum
manifold is topologically nontrivial.  The arguments which say that in
a theory which admits defects, such defects will inevitably form, are
applicable both in cosmology and in condensed matter physics.
Different, however, is the defect dynamics.  The motion of defects in
condensed matter systems is friction-dominated, whereas the defects in
cosmology obey relativistic equations, second order in time
derivatives, since they come from a relativistic field theory.

Turning to a classification of
topological defects, we consider theories with an $n$-component order
parameter $\varphi$ and with a potential energy function (free energy
density) of the form (\ref{pot}) with
\be
\varphi^2 = \sum\limits^n_{i = 1} \, \varphi^2_i \, .  
\ee

There are various types of local and global topological defects
(regions of trapped energy density) depending on the number $n$ of components
of $\varphi$.  The more rigorous mathematical definition refers to the homotopy
of ${\cal M}$.  The words ``local" and ``global" refer to whether the symmetry
which is broken is a gauge or global symmetry.  In the case of local
symmetries, the topological defects have a well defined core outside of which
$\varphi$ contains no energy density in spite of nonvanishing gradients
$\nabla \varphi$:  the gauge fields $A_\mu$ can absorb the gradients,
{\it i.e.,} $D_\mu \varphi = 0$ when $\partial_\mu \varphi \neq 0$,
where the covariant derivative $D_\mu$ is defined by
\be
D_\mu = \partial_\mu + ie \, A_\mu \, ,  
\ee
$e$ being the gauge coupling constant.
Global topological defects, however, have long range density fields and
forces.
 
Table 1 contains a list of topological defects with their topological
characteristic.  A ``v" marks acceptable theories, a ``x" theories which are
in conflict with observations (for $\eta \sim 10^{16}$ GeV).
 
\begin{table}[htb]
\begin{center}
\caption{Classification of cosmologically allowed (v) and forbidden (x) defects.}
\begin{tabular}{llll}
\hline
Defect name & n & Local defect & Global defect\\
\hline
Domain wall & 1 & x & x\\
Cosmic string & 2 & v & v\\
Monopole & 3 & x & v\\
Texture & 4 & - & v\\
\hline
\end{tabular}
\end{center}
\end{table}

We now describe examples of domain walls, cosmic strings, monopoles and
textures, focussing on configurations with maximal symmetry.

{\it Domain walls} arise in theories with a single real order
parameter and free energy density given by (\ref{pot}).  The vacuum manifold
of this model consists of two points
\be
{\cal M} = \{ \pm \eta \}  
\ee
and hence has nontrivial zeroth homotopy group:
\be
\Pi_0 ({\cal M}) \neq 1  \, \ .  
\ee
 
To construct a domain wall configuration with planar symmetry (without
loss of generality the $y-z$ plane can be taken to be the plane of
symmetry), assume that
\begin{eqnarray}
\varphi (x) \simeq \eta \>\>\> & x \gg \eta^{-1} \\
\varphi (x) \simeq - \eta \>\>\> & x \ll - \eta^{-1} 
\end{eqnarray}
By continuity of $\varphi$, there must be an intermediate value of $x$
with $\varphi (x) = 0$.   
The set of points with $\varphi = 0$ constitute the center of the
domain wall.  Physically, the wall is a thin sheet of trapped energy
density.  The width $w$ of the sheet is given by the balance of potential
energy and tension energy.  Assuming that the spatial gradients are
spread out over the thickness $w$ we obtain
\be
w V (0) = w \lambda \eta^4 \sim {1\over w} \, \eta^2  
\ee
and thus
\be \label{width}
w \sim \lambda^{-1/2} \eta^{-1} \, .  
\ee
   
A theory with a complex order parameter $(n = 2)$ admits
{\it cosmic strings}.   In this case the vacuum manifold of the
model is
\be
{\cal M} = S^1 \, ,  
\ee
which has nonvanishing first homotopy group:
\be
\Pi_1 ({\cal M}) = Z \neq 1 \, .  
\ee
A cosmic string is a line of trapped energy density which arises
whenever the field $\varphi (x)$ circles ${\cal M}$ along a closed path
in space ({\it e.g.}, along a circle).  In this case, continuity of
$\varphi$ implies that there must be a point with $\varphi = 0$ on any
disk whose boundary is the closed path.  The points on different sheets
connect up to form a line over-density of field energy, a cosmic string.  

To construct a field configuration with a string along the $z$ axis \cite{NO},
take $\varphi (x)$ to cover ${\cal M}$ along a circle with radius $r$
about the point $(x,y) = (0,0)$:
\be \label{string}
\varphi (r, \vartheta ) \simeq \eta e^{i \vartheta} \, , \, r \gg
\eta^{-1} \, .  
\ee
This configuration has winding number 1, {\it i.e.}, it covers ${\cal
M}$ exactly once.  Maintaining cylindrical symmetry, we can extend
(\ref{string}) to arbitrary $r$
\be
\varphi (r, \, \vartheta) = f (r) e^{i \vartheta} \, ,  
\ee
where $f (0) = 0$ and $f (r)$ tends to $\eta$ for large $r$.  The
width $w$ can again be found by balancing potential and tension
energy.  The result is identical to the result (\ref{width}) for domain walls.

For local cosmic strings, {\it i.e.}, strings arising due to the
spontaneous breaking of a gauge symmetry, the energy density decays
exponentially for $r \gg \eta^{-1}$.  In this case, the energy $\mu$
per unit length of a string is finite and depends only on the symmetry
breaking scale $\eta$
\be
\mu \sim \eta^2  
\ee
(independent of the coupling $\lambda$).  The value of $\mu$ is the
only free parameter in a cosmic string model.

If the order parameter of the model has three components $(n = 3)$,
then {\it monopoles} result as topological defects.  The vacuum
manifold is
\be
{\cal M} = S^2  
\ee
and has topology given by
\be
\Pi_2 ({\cal M}) \neq 1 \, .  
\ee
Given a sphere $S$ in space, it is possible that $\varphi$ takes on
values in ${\cal M}$ everywhere on $S$, and that it covers ${\cal M}$
once.  By continuity, there must be a point in space in the interior
of $S$ with $\varphi = 0$.  This is the center of a point-like defect,
the monopole.

To construct a spherically symmetric monopole with the origin as its
center, consider a field configuration $\varphi$ which defines a map
from physical space to field space such that all spheres $S_r$ in
space of radius $r \gg \eta^{-1}$ about the origin are mapped onto
${\cal M}$:
\begin{eqnarray}
\varphi: \> & S_r \longrightarrow {\cal M} \\
& (r, \vartheta,\varphi)  \longrightarrow (\vartheta,
\varphi)  \, .  
\end{eqnarray}
This configuration defines a winding number one magnetic monopole.

Domain walls, cosmic strings and monopoles are examples of
{\it topological} defects.  A topological defect has a
well-defined
core, a region in space with $\varphi \notin {\cal M}$ and hence $V
(\varphi) > 0$.  There is an associated winding number which is
quantized, {\it i.e.}, it can take on only integer values.  Since the
winding number can only change continuously, it must be conserved, and
hence topological defects are stable.  Furthermore, topological
defects exist for theories with global and local symmetries.

Now, let us consider a theory with a four-component order parameter
({\it i.e.,} $n = 4$), and a potential given by (\ref{pot}).  In this case,
the vacuum manifold is
\be
{\cal M} = S^3  
\ee
and the associated topology is given by
\be
\Pi_3 ({\cal M}) \neq 1 \, .  
\ee
The corresponding defects are called {\it textures} \cite{Davis,Turok}. 

Textures, however, are quite different than the previously discussed topological defects.
The texture construction will render this manifest.  To construct a
radially symmetric texture, we give a field configuration $\varphi (x)$ which
maps physical space onto ${\cal M}$.  The origin 0 in space (an arbitrary point
which will be the center of the texture) is mapped onto the north pole $N$ of
${\cal M}$.  Spheres in space of radius $r$ surrounding 0 are mapped onto spheres in ${\cal M}$ surrounding $N$, with distance from $N$ increasing as $r$ increases.  In
particular, some sphere with radius $r_c (t)$ is mapped onto the equator
sphere of ${\cal M}$.  The distance $r_c (t)$ can be defined as the radius of
the texture.  Inside this sphere, $\varphi (x)$ covers half the vacuum
manifold.
Finally, the sphere at infinity is mapped onto the south pole of ${\cal M}$.
The configuration $\varphi ({\it x})$ can be parameterized
by \cite{Turok}
\be
\varphi (x,y,z) = \left(\cos \chi (r), \> \sin \chi (r) {x\over r}, \>
\sin \chi (r) {y\over r}, \> \sin \chi (r) {z\over r} \right)  
\ee
in terms of a function $\chi (r)$ with $\chi (0) = 0$ and $\chi (\infty) =
\pi$.  Note that at all points in space, $\varphi ({\it x})$ lies in
${\cal M}$.  There is no defect core.  All the energy is in spatial gradient (and possibly kinetic) terms.
 
In a cosmological context, there is infinite energy available in an infinite
space.  Hence, it is not necessary that $\chi (r) \rightarrow \pi$ as $r
\rightarrow \infty$.  We can have
\be
\chi (r) \rightarrow \chi_{\rm max} < \pi \>\> {\rm as} \>\> r \rightarrow
\infty \, .  
\ee
In this case, only a fraction
\be
n_W = {{\chi_{\rm max}} \over {\pi}} - {{\sin (2 \chi_{\rm max})} \over{2 \pi}}
\ee
of the vacuum manifold is covered:  the winding number $n_W$ is not quantized.
This is a reflection of the fact that whereas topologically nontrivial maps
from $S^3$ to $S^3$ exist, all maps from $R^3$ to $S^3$ can be deformed to
the trivial map.
 
Textures in $R^3$ are unstable.  For the configuration described above, the
instability means that $r_c (t) \rightarrow 0$ as $t$ increases: the texture
collapses.  When $r_c (t)$ is microscopical, there will be sufficient energy
inside the core to cause $\varphi (0)$ to leave ${\cal M}$, pass through 0 and
equilibrate at $\chi (0) = \pi$: the texture unwinds.
 
A further difference compared to topological defects: textures are relevant
only for theories with global symmetry.  Since all the energy is in spatial
gradients, for a local theory the gauge fields can re-orient themselves such as
to cancel the energy:
\be
D_\mu \varphi = 0 \, .  
\ee 

Therefore, it is reasonable to regard textures as an example of a new class of
defects, {\it semi-topological defects}.  In contrast to topological
defects, there is no core, and $\varphi ({\it x}) \in {\cal M}$ for all
${\it x}$.  In particular, there is no potential energy.  In addition,
the winding number is not quantized, and hence the defects are unstable.  Finally, they exist as long-lived coherent configurations only in theories with a global internal symmetry.
 
The Kibble mechanism \cite{Kibble} ensures that in theories which admit
topological or semi-topological defects, such defects will be produced
during a phase transition in the very early Universe.
At high temperatures $T \gg T_c$, the symmetry is unbroken. The ground state of the finite temperature effective potential \cite{KL,DJ} (see e.g. \cite{RB85} for an introductory review) is the symmetric configuration $\varphi = 0$.  
Once the Universe cools below the temperature $T_c$, the symmetry is broken, and $\varphi(x)$ rolls into the zero temperature vacuum manifold ${\cal M}$. However, by causality there can be no correlation between the specific points in ${\cal M}$ which are taken on on scales larger than the correlation length $\xi (t)$. In a relativistic theory the causality bound on $\xi$ is
\be
\xi (t) < t \, ,  
\ee
where $t$ is the causal horizon. This leads to the formation of defects with a mean separation of $\xi (t)$.

The correlation length $\xi (t)$ can be determined by equating the free energy gained by symmetry breaking (a volume effect) with the gradient energy lost (a surface effect).  As expected, $\xi (T)$ diverges at $T_c$. Very close to $T_c$, the thermal energy $T$ is larger than the volume energy gain $E_{corr}$ in a
correlation volume. Hence, these domains are unstable to thermal fluctuations.
As $T$ decreases, the thermal energy decreases more rapidly than $E_{corr}$.
Below the Ginsburg temperature $T_G$, there
is insufficient thermal energy to excite a correlation volume into the
state $\varphi = 0$.  Domains of size
\be \label{GBL}
\xi (t_G) \sim \lambda^{-1} \eta^{-1}  
\ee
freeze out \cite{Kibble,Kibble2}.  The boundaries between these domains become
topological defects. An improved version of this argument has been given by Zurek \cite{Zurek}. 

We conclude that in a theory in which a symmetry breaking phase
transition satisfies the topological criteria for the existence of a
given type of defect, a network of such defects will form during the
phase transition and will freeze out at the Ginsburg temperature.  The
correlation length is initially given by (\ref{GBL}), if the field
$\varphi$ is in thermal equilibrium before the transition.
Independent of this last assumption, the causality bound implies that
$\xi (t_G) < t_G$.

For times $t > t_G$ the evolution of the network of defects may be
complicated (as for cosmic strings) or trivial (as for textures).  In
any case (see the caveats of Refs. \cite{caveat1,caveat2}), the causality bound
persists at late times and states that even at late times, the mean
separation and length scale of defects is bounded by $\xi (t) \leq t$.

Applied to cosmic strings, the Kibble mechanism implies that at the
time of the phase transition, a network of cosmic strings with typical
step length $\xi (t_G)$ will form.  According to numerical
simulations \cite{VV1}, about 80\% of the initial energy is in infinite
strings (strings with curvature radius larger than the Hubble radius) and 20\% in closed loops.

The evolution of the cosmic string network for $t > t_G$ is
complicated.  The key processes are loop production
by intersections of infinite strings  and loop shrinking
by gravitational radiation.  These two processes combine to create a
mechanism by which the infinite string network loses energy (and
length as measured in comoving coordinates). The dynamics of a string in an expanding Universe changes as the curvature radius increases. If the curvature radius is smaller than the Hubble radius $H^{-1}(t)$, the dynamics is dominated by the acceleration term and the string will oscillate at relativistic speeds. However, if the curvature radius exceeds the Hubble radius, the strings are frozen in by the Hubble damping term in the string equation of motion. Therefore, if the string correlation length is smaller than the Hubble radius, string intersections will be frequent, the network will loose a lot of energy to loops and $\xi(t)$ will increase faster than $t$. However, if at any time $\xi(t)$ exceeds the Hubble radius, the strings will be effectively frozen in comoving coordinates and hence in a standard radiation or matter dominated cosmology the Hubble radius will catch up to $\xi(t)$. According to this argument \cite{Vil85}, at sufficiently late times the correlation leng!
th of the string network will always be proportional to its causality limit
\be \label{scaling}
\xi (t) \sim t \, .  
\ee
Hence, the energy density $\rho_\infty (t)$ in long strings is a fixed
fraction of the background energy density $\rho_c (t)$
\be
\rho_\infty (t) \sim \mu \xi (t)^{-2} \sim \mu t^{-2}  
\ee
or
\be
{\rho_\infty (t)\over{\rho_c (t)}} \sim G \mu \, .  
\ee

We conclude that the cosmic string network approaches a ``scaling
solution" in which the statistical properties of the
network are time independent if all distances are scaled to the
Hubble radius.

Applied to textures (see e.g. \cite{Turok2} for a review), the Kibble mechanism implies that on all
scales $r \geq t_G$, field configurations with winding number $n_W
\geq n_{cr}$ are frozen in with a probability $p (n_{cr})$ per volume
$r^3$.  The critical winding number $n_{cr}$ is defined as the winding
number above which field configurations collapse and below which they
expand.  Only collapsing configurations form clumps of energy which
can accrete matter. For spherically symmetric textures, 
the critical winding $n_{cr}$ is slightly larger than 0.5 \cite{ncr}.   

For $t > t_G$, any configuration on scale $\sim t$ with winding number
$n_W \ge n_{cr}$ begins to collapse (if the scale of the texture is larger than $t$, the Hubble damping
term dominates over the spatial gradient forces, and the field
configuration is frozen in comoving coordinates).  After unwinding,
$\varphi ({\it x})$ is homogeneous inside the horizon.

The texture model thus also leads to a scaling solution: at all times
$t > t_G$ there is the same probability that a texture configuration
of scale $t$ will enter the horizon, become dynamical and collapse
on a typical time scale $t$.

\section{Defects and Structure Formation: An Overview}

Topological defects are regions in space with trapped energy density.
By Newtonian gravity, these defects can act as seeds about which the
matter in the Universe clusters, and hence they play a very important
role in cosmology.

As indicated in Table 1, theories with domain walls or with local
monopoles are ruled out, and those with only local textures do not give
rise to a structure formation model.  As mentioned earlier, theories with
domain walls are
ruled out since a single wall stretching across the present Universe
would overclose it.  Local monopoles are also problematic since they
do not interact and come to dominate the energy density of the
Universe.  Local textures do not exist as coherent structures with
nonvanishing gradient energy since the gauge fields can always
compensate scalar field gradients.
\par
Let us demonstrate explicitly why stable domain walls are a
cosmological disaster \cite{Zel74}.  If domain walls form during a phase transition
in the early Universe, it follows by causality (see however the caveats
of Refs. \cite{caveat1,caveat2}) that even today there will be at least one wall
per Hubble volume.  Assuming one wall per Hubble volume, the energy
density $\rho_{DW}$ of matter in domain walls is
\be
\rho_{DW} (t) \sim \eta^3 t^{-1} \, ,  
\ee
whereas the critical density $\rho_c$ is
\be
\rho_c = H^2 \, {3\over{8 \pi G}} \sim m^2_{p\ell} \, t^{-2} \, .
 \ee
Hence, for $\eta \sim 10^{16}$ GeV the ratio evaluated at the present time $t_0$ is
\be
{\rho_{DW}\over \rho_c} \, (t_0) \sim \, \left({\eta\over{m_{p\ell}}}
\right)^2 \, (\eta t_0) \sim 10^{52} \, .  
\ee

The above argument depends in an essential way on the dimension of the
defect.  One cosmic string per Hubble volume leads to an energy
density $\rho_{cs}$ in string
\be
\rho_{cs} \sim \eta^2 \, t^{-2} \, .  
\ee
As we have seen above, the cosmic string network approaches a scaling distribution (\ref{scaling}). Hence, cosmic strings do not lead to
cosmological problems.  On the contrary, since for GUT models with
$\eta \sim 10^{16}$ GeV
\be
{\rho_{cs}\over \rho_c} \sim \, \left({\eta\over m_{p \ell}} \right)^2
\sim 10^{-6} \, ,  
\ee
cosmic strings in such theories could provide the seed perturbations
responsible for structure formation.

Theories with local monopoles are ruled out on cosmological
grounds \cite{Zel78} (see again the caveats of Refs. \cite{caveat1,caveat2}) for
rather different reasons.  Since there are no long range forces
between local monopoles, their number density in comoving coordinates
does not decrease.  Since their contribution to the energy density
scales as $a^{-3} (t)$, they will come to dominate the mass of the
Universe, provided $\eta$ is sufficiently large.

Theories with global monopoles \cite{BV89,RB90} are not ruled out, since
there are long range forces between monopoles which lead to a
``scaling solution" with a fixed number of monopoles per Hubble
volume.

Let us now briefly discuss some of the basic mechanisms of the cosmic string scenario of structure formation. The implementation of the mechanisms in concrete models will be covered in the following sections.

The starting point of the structure formation scenario in the cosmic
string theory is the scaling solution for the cosmic string network,
according to which at all times $t$ (in particular at $t_{eq}$, the
time when perturbations can start to grow) there will be a few long
strings crossing each Hubble volume, plus a distribution of loops of
radius $R \ll t$. 

The cosmic string model admits three mechanisms for structure
formation:  loops, filaments, and wakes.  Cosmic string loops have the same
time averaged field as a point source with mass \cite{Turok84}
\be
M (R) = \beta R \mu \, ,  
\ee
$R$ being the loop radius and $\beta \sim 2 \pi$.  Hence, loops will be seeds
for spherical accretion of dust and radiation.

For loops with $R \leq t_{eq}$, growth of perturbations in a model
dominated by cold dark matter starts at $t_{eq}$.  Hence, the mass at
the present time will be
\be
M (R, \, t_0) = z (t_{eq}) \beta \, R \mu \, .  
\ee

In the original cosmic string model \cite{ZelVil,TB86} it was assumed
that loops dominate over wakes.  In this case, the theory could be
normalized ({\it i.e.}, $\mu$ could be determined) by demanding that loops
with the mean separation of clusters $d_{cl}$ accrete the correct mass, {\it i.e.}, that
\be
M (R (d_{cl}), t_0) = 10^{14} M_{\odot} \, .  
\ee
This condition yields \cite{TB86}
\be
\mu \simeq 10^{32} {\rm GeV}^2  
\ee
Thus, if cosmic strings are to be relevant for structure formation,
they must arise due to a symmetry breaking at an energy scale $\eta
\simeq 10^{16}$GeV.  This scale happens to be the scale of unification (GUT)
of weak, strong and electromagnetic interactions.  It is tantalizing
to speculate that cosmology is telling us that there indeed was new
physics at the GUT scale.

\begin{figure}[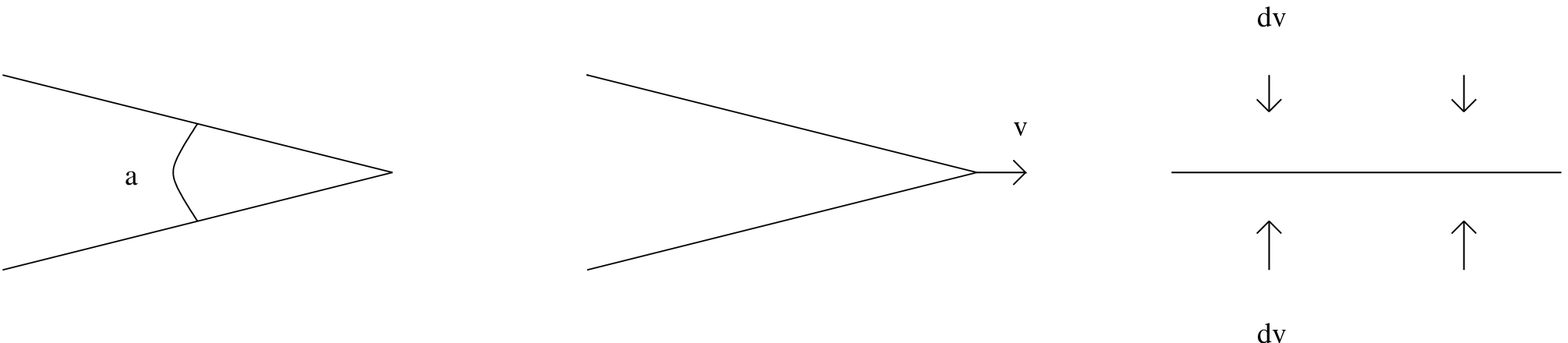]
\begin{center}
\leavevmode
\epsfysize=5cm \epsfxsize=12cm \epsfbox{kish2fig1.eps}
\end{center}
\caption{Sketch of the mechanism by which a long
straight cosmic string $S$ moving with velocity $v$ in transverse
direction through a plasma induces a velocity perturbation $\Delta v (dv)$
towards the wake. Shown on the left is the deficit angle $a$, in the
center is a sketch of the string moving in the plasma, and on the
right is the sketch of how the plasma moves towards the wake with velocity $dv$ in the frame in which the string is at rest.}
\label{fig0}
\end{figure}

The second mechanism involves long strings moving with relativistic
speed in their normal plane which give rise to
velocity perturbations in their wake \cite{SV84}.  The mechanism is illustrated in Fig. 1: space normal to the string is a cone with deficit angle \cite{V81}
\be \label{deficit}
\alpha = 8 \pi G \mu \, .  
\ee
If the string is moving with normal velocity $v$ through a bath of dark
matter, a velocity perturbation
\be
\delta v = 4 \pi G \mu v \gamma  
\ee
[with $\gamma = (1 - v^2)^{-1/2}$] towards the plane behind the string
results.  At times after $t_{eq}$, this induces planar over-densities,
the most
prominent ({\it i.e.}, thickest at the present time) and numerous of which were
created at $t_{eq}$, the time of equal matter and
radiation \cite{TV86,SVBST,BPS}.  The
corresponding planar dimensions are (in comoving coordinates)
\be
t_{eq} z (t_{eq}) \times t_{eq} z (t_{eq}) v \sim (40 \times 40 v) \,
{\rm Mpc}^2
\, .  
\ee

The thickness $d$ of these wakes can be calculated using the
Zel'dovich approximation \cite{BPS}.  The result is
\be
d \simeq G \mu v \gamma (v) z (t_{eq})^2 \, t_{eq} \simeq 4 v \, {\rm
Mpc} \, .  
\ee
 
Wakes arise if there is little small scale structure on the string.
In this case, the string tension equals the mass density, the string
moves at relativistic speeds, and there is no local gravitational
attraction towards the string.

In contrast, if there is small scale structure on strings,
then the string tension $T$ is smaller \cite{BC90} than the mass per unit
length $\mu$ and the metric of a string in $z$ direction becomes \cite{VV91}
\be \label{metric2}
ds^2 = (1 + h_{00}) (dt^2 - dz^2 - dr^2 - (1 - 8G \mu) r^2 dy^2 )
\ee
with
\be
h_{00} = 4G (\mu - T) \ln \, {r\over r_0} \, ,  
\ee
$r_0$ being the string width.  Since $h_{00}$ does not vanish, there
is a gravitational force towards the string which gives rise to
cylindrical accretion, thus producing filaments.

As is evident from the last term in the metric (\ref{metric2}), space
perpendicular to the string remains conical, with deficit angle given
by (\ref{deficit}).  However, since the string is no longer relativistic, the
transverse velocities $v$ of the string network are expected to be
smaller, and hence the induced wakes will be shorter and thinner.

Which of the mechanisms -- filaments or wakes -- dominates is
determined by the competition between the velocity induced by $h_{00}$
and the velocity perturbation of the wake.  The total velocity
is \cite{VV91}
\be
u = - {2 \pi G (\mu - T)\over{v \gamma (v)}} - 4 \pi G \mu v \gamma
(v) \, ,  
\ee
the first term giving filaments, the second producing wakes.  Hence,
for small $v$ the former will dominate, for large $v$ the latter.

By the same argument as for wakes, the most numerous and prominent
filaments will have the distinguished scale
\be
t_{eq} z (t_{eq}) \times d_f \times d_f  
\ee
where $d_f$ can be calculated using the Zel'dovich approximation \cite{AB95}.

The cosmic string model predicts a scale-invariant spectrum of density
perturbations, exactly like inflationary Universe models but for a
rather different reason.  Consider the {\it r.m.s.} mass fluctuations
on a length scale $2 \pi k^{-1}$ at the time $t_H (k)$ when this scale
enters the Hubble radius $H^{-1}(t)$.  From the cosmic string scaling solution it follows that a fixed ({\it i.e.}, $t_H (k)$ independent) number
$\tilde v$ of strings of length of the order $t_H (k)$ contribute to
the mass excess $\delta M (k, \, t_H (k))$.  Thus
\be
{\delta M\over M} \, (k, \, t_H (k)) \sim \, {\tilde v \mu t_H
(k)\over{G^{-1} t^{-2}_H (k) t^3_H (k)}} \sim \tilde v \, G \mu \, .
\ee
Note that the above argument predicting a scale invariant spectrum
will hold for all topological defect models which have a scaling
solution, in particular also for global monopoles and textures.

The amplitude of the {\it r.m.s.} mass fluctuations (equivalently: of
the power spectrum) can be used to normalize $G \mu$.  Since today on
galaxy cluster scales
\be
{\delta M\over M} (k, \, t_0) \sim 1 \, ,  
\ee
the growth rate of fluctuations linear in $a(t)$ yields
\be
{\delta M\over M} \, (k, \, t_{eq}) \sim 10^{-4} \, ,  
\ee
and therefore, using $\tilde v \sim 10$,
\be
G \mu \sim 10^{-5} \, .  
\ee

In contrast to the situation in inflationary Universe
models, hot dark matter (HDM) is not from the outset ruled out as a dark matter candidate. As non-adiabatic seeds, cosmic string loops survive free streaming and can
generate nonlinear structures on galactic scales, as discussed in
detail in Refs. \cite{BKST,BKT}.  Accretion of hot dark matter by a string wake
was studied in Ref. \cite{BPS}. In this case, nonlinear perturbations
develop only late.  At some time $t_{nl}$, all scales up to a distance
$q_{\rm max}$ from the wake center go nonlinear.  Here
\be
q_{\rm max} \sim G \mu v \gamma (v) z (t_{eq})^2 t_{eq} \sim 4 v \,
{\rm Mpc} \, ,  
\ee
is the  comoving thickness of the wake at $t_{nl}$.  Demanding
that $t_{nl}$ corresponds to a redshift greater than 1 leads to the
constraint
\be
G \mu > 5 \cdot 10^{-7} \, .  
\ee
Note that in a cosmic string and hot dark matter model, wakes form nonlinear structures only very recently. Accretion onto loops \cite{MB96} and onto
filaments \cite{ZLB} provide two mechanisms which may lead to high redshift objects such as quasars and high redshift galaxies.

\section{Introduction to high precision calculations with defects}

The last decade has
witnessed unprecedented progress in mapping the
cosmic microwave background (CMB) 
temperature anisotropy
and the large scale structure (LSS) of the Universe. 
The prospect of fast improving data has
forced theorists to 
new standards of precision in computing observable quantities.
The new standards have been met in theories based
on cosmic inflation\cite{hsselj,hw}. 
Topological defect scenarios \cite{ShellVil,HK95}
have been more challenging. However, 
recently there have been a number of computational breakthroughs in defect
theories, partly related to improvements in computer technology.
Most strikingly, the method described in \cite{pst} showed how
one could glean from defect simulations all the information required
to compute accurately CMB and LSS power spectra.
In this method the simulations are used uniquely for  evaluating
the two point functions (known as unequal time
correlators, or UETCs) of the defects' stress-energy tensor. 
UETCs are all that is required for computing CMB and LSS power spectra. 
Furthermore, they are constrained by requirements of self-similarity 
(or scaling) and
causality, which enable us to radically extend the dynamical 
range of simulations, a fact 
central to the success of the method.

This method was applied to theories based on global
symmetries. In recent work \cite{chm,chm1} we have shown how the same
method could be applied to local cosmic strings (see also \cite{steb,abr}). 
In the next few Sections we shall describe in detail the simulation
and  measurement of UETCs which led to the work in \cite{chm,chm1};
as well as present the analytical tools used in this enterprise. 
The formalism we had to use is unfortunately more complicated than
\cite{pst}. Local strings have an extra complication over global defects, which stems 
from the fact that we are unable to simulate the underlying field theory. 
Instead, we approximate the true dynamics with line-like relativistic strings. 
This is thought to be reasonable for the large scale properties of the 
stress-energy tensor, but we do not have a good understanding of how 
the string network loses energy in order to maintain scaling.  

This leads
to two problems. Firstly one is forced
to make assumptions about which cosmological fluids pick up this deficit. 
It is often assumed that all the strings' energy and momentum is radiated 
into gravitational waves, approximated by a relativistic fluid. 
This is by no means certain, and it may well be that the energy and momentum
is transferred to particles \cite{VinAntHin98}, 
and hence to the baryon, photon and CDM 
components.  

Secondly, it is not enough to find
correlators for a reduced number of stress energy tensor
components (two scalar, two vector, two tensor), and then 
find the others by means of energy conservation. If energy 
conservation can  be used then one needs to compute 3 scalar,
1 vector, and 1 tensor UETCs. If one is not allowed
to make use of energy conservation, one has to compute 10
scalar, 3 vector, and 1 tensor UETC. 

In the following sections we first give a qualitative description
of the technical novelties introduced in defect scenarios. We describe
defects as active incoherent perturbations. We then describe 
a set of tools with which we can perform high accuracy calculations
of structure power spectra in these scenarios. 

\section{Defects as active, incoherent perturbations}

We first focus on the basic assumptions of
inflationary and defect theories and  isolate  the most striking
contrasting properties. 
We define the concepts of active and passive perturbations,
and of coherent and incoherent perturbations. In terms of 
these concepts inflationary perturbations are 
passive coherent perturbations.
Defect perturbations are active perturbations more or less
incoherent depending on the defect \cite{inc}.

\subsection{Active and passive perturbations, and their different
perceptions of causality and scaling}

The way in which inflationary and defect perturbations come about
is radically different. Inflationary fluctuations 
were produced at a remote epoch, and were 
driven far outside the Hubble radius by  inflation.  The
evolution of these fluctuations is linear (until 
gravitational collapse becomes non-linear at late times), and we call
these fluctuations ``passive''.  Also, because
all scales observed today have been in causal contact since the onset
of inflation, causality does not strongly constrain the fluctuations
which result. In contrast, defect fluctuations are continuously seeded by
defect evolution, which is a non-linear process.
We therefore say these are ``active'' perturbations.  Also, the
constraints imposed by causality on defect formation  and evolution 
are much greater than  those placed on inflationary perturbations.

\subsubsection{Active and passive scaling}

The notion of scale invariance has different implications
in these two types of theory. For instance, a scale invariant gauge-invariant
potential $\Phi$ with dimensions $L^{3/2}$ has a power spectrum 
$$P(\Phi)=\langle |\Phi_{\bf k}|^2\rangle\propto k^{-3}$$ 
in passive theories (the Harrison-Zel'dovich spectrum). 
This results from the fact that
the only variable available is $k$, and so the only spectrum one
can write down which has the right dimensions and does not have a 
scale is the Harrison-Zel'dovich spectrum. 
The situation is different
for active theories, since time is now a variable.
The  most general counterpart to the Harrison-Zel'dovich spectrum is  
\begin{equation}\label{scale}
P(\Phi) =  \eta^3F_{\Phi}(k\eta)
\end{equation}
where $F_{\Phi}$
is, to begin with, an arbitrary function of $x=k\eta$. All other
variables may be written as a product of a power of $\eta$, ensuring
the right dimensions, and an arbitrary function of $x$. 
Inspecting all equations it can be checked that it is possible to do
this consistently for all variables. All equations respect scaling
in the active sense.

\subsubsection{Causality constraints on active perturbations}\label{causal}

Moreover, active perturbations are constrained by causality, in the form of 
integral constraints \cite{trasch12,james}. These consist of 
energy and momentum conservation laws for fluctuations
in an expanding Universe. The integral constraints can be used to 
find the low $k$ behaviour of the power spectrum of the perturbations, 
assuming their causal generation and evolution
\cite{traschk4}. Typically, it is found that the causal creation
and evolution of defects requires that their energy $\rho^s$ and scalar
velocity $v^s$
be white noise at low $k$, but that the total energy power spectrum 
of the fluctuations is required to go like $k^4$. To reconcile these two facts
one is forced to consider the compensation. 
This is an under-density in the matter-radiation  energy density
with a white noise low $k$ tail, correlated with the defect network
so as to cancel the defects' white-noise tail. When one combines the
defects energy with the compensation density, one finds that the 
gravitational potentials they generate also have to be
white noise at large scales \cite{inc}. Typically the scaling function
$ F_{\Phi}(k\eta)$ will start as a constant and decay as a power law
for $x=k\eta>x_c$. The value $x_c$ is a sort of coherence wavenumber
of the defect. The larger it is the smaller the defect is. For instance
$x_c\approx 12$ for cosmic strings (thin, tiny objects), whereas
$x_c\approx 5.5$ for textures (round, fat, big things). Sophisticated
work on causality \cite{james} has shed light on how small $x_c$
may be before violating causality. The limiting lower bound $x_c
\approx 2.7$  has been suggested.

Although we will not here have a chance to dwell on technicalities,
it should be stated that the rather general discussion presented above
is enough to determine the general form of the potentials for active 
perturbations. This has been here encoded in the single parameter $x_c$.
We shall see that $x_c$ will determine the Doppler peak position
for active perturbations. Doppler peaks are driven by the gravitational
potential, so it should not be surprising that the defect length scale
propagates into its potential, and from that  into the position of the 
Doppler peaks.

\subsection{Coherent and incoherent perturbations}

Active perturbations may also differ from
inflation in the way ``chance'' comes into the theory. 
Randomness occurs in inflation only when the initial 
conditions are set up. Time evolution is linear and
deterministic, and may be found by
evolving all variables from an
initial value equal to the square root of 
their initial variances. By squaring the
result one obtains the variances of the variables at any time.
Formally, this results from unequal time correlators of the form
\begin{equation}\label{2cori}
{\langle\Phi({\bf k},\eta)\Phi({\bf k'},\eta ')\rangle}=
\delta({\bf k}-{\bf k'})\sigma({\Phi}(k,\eta))\sigma
({\Phi}(k,\eta')),
\end{equation}
where $\sigma$ denotes the square root of the power spectrum $P$.
In defect models however, randomness may intervene in the time
evolution as well  as the initial conditions. 
Although deterministic in principle, 
the defect network evolves as a result of a 
complicated non-linear process.
If there is strong non-linearity, a given mode will be ``driven'' 
by interactions with the other modes in a way which will force
all different-time correlators to zero on a time scale
characterized by the ``coherence time'' $\theta_c(k,\eta)$.
Physically this means that one has to perform a new ``random'' draw 
after each coherence time in order to
construct a defect history \cite{inc}. 
The counterpart to (\ref{2cori}) for incoherent perturbations is
\begin{equation}\label{pr0}
{\langle\Phi({\bf k},\eta)\Phi({\bf k'},\eta ')\rangle}=
\delta({\bf k}-{\bf k'}) P({\Phi}(k,\eta),\eta'-\eta)\; .
\end{equation}
For $|\eta'-\eta| \equiv |\Delta\eta|> \theta_c(k,\eta)$
we have $P({\Phi}(k,\eta),\Delta\eta)=0$. For $\Delta\eta=0$,
we recover the power spectrum $P({\Phi}(k,\eta),0)=P({\Phi}(k,\eta))$.

We shall label as coherent and incoherent
(\ref{2cori}) and (\ref{pr0})  respectively. 
This feature does not affect the position of the Doppler peaks,
but it does affect the structure of secondary oscillations.
An incoherent potential will drive the CMB oscillator incoherently,
and therefore it may happen that the secondary oscillations get
washed out as a result of incoherence.

\section{Tool 1: Scalar, vector, and tensor decomposition}\label{svt}
Having identified the main qualitative novelties in defecy calculations,
we now proceed to present the set of tools required for performing
high accuracy calculations in these scenarios. We start with the decomposition
into scalar, vector, and tensor components. 
Let $\Theta_{\mu\nu}({\bf x})$ be the defect stress-energy tensor. 
We may Fourier analyze it 
\begin{equation}
\Theta_{\mu\nu}({\bf x})={\int d^3k}\Theta_{\mu\nu}({\bf k})
e^{i{\bf k}\cdot {\bf x}}
\end{equation}
and decompose its Fourier components as:
\begin{eqnarray}
\Theta_{00}&=&\rho^d\\
\Theta_{0i}&=&i{\hat k_i}v^d+\omega^d_i\\
\Theta_{ij}&=&p^d\delta_{ij}+
{\left({\hat k_i}{\hat k_j}-{1\over 3}\delta_{ij}\right)}
\Pi^S+\nonumber\\
&&i{\left({\hat k_i}\Pi_j^V+{\hat k_j}\Pi_i^V\right)}
+\Pi_{ij}^T
\end{eqnarray}
with ${\hat k^i}\omega^d_i=0$, ${\hat k^i}\Pi^V_i=0$,
${\hat k^i}\Pi_{ij}^T=0$, and $\Pi^{Ti}_{i}=0$.
The variables $\{\rho^d, v^d, p^d, \Pi^S\}$
are the scalars, $\{ \omega^d_i, \Pi^V_i\}$ the vectors,
and $\Pi^T_{ij}$ the tensors. The decomposition
can be inverted by means of
\begin{eqnarray}
v^d&=&-i{\hat k^i}\Theta_{0i}\\
\omega^d_i&=&(\delta_i^j-{\hat k^i}{\hat k_j})\Theta_{0j}
\end{eqnarray}
and
\begin{eqnarray}
p^s&=&{1\over 3}\Theta^i_i\\
\Pi^S&=&{3\over 2}({\hat k^i}{\hat k^j}-{1\over 3}
\delta^{ij})\Theta_{ij}\\
\Pi^V_i&=&-i{\left({\hat k^j}\delta^l_i-{\hat k_i}{\hat k^j}
{\hat k^l}\right)}\Theta_{lj}\\
\Pi^T_{ij}&=&( \delta^l_i\delta^m_j-{1\over 2}\delta_{ij}\delta^{lm}
+{1\over 2} {\hat k_i}{\hat k_j}{\hat k^l}{\hat k^m}
+{1\over 2} ({\hat k^l}{\hat k^m}\delta_{ij}+\nonumber\\
&&{\hat k_i}{\hat k_j}\delta^{lm})- ({\hat k_i}{\hat k^l}
\delta^m_j+{\hat k_j}{\hat k^l}\delta^m_i) )\Theta_{lm}
\end{eqnarray}
although simpler recursive formulae may be written.

This decomposition and choice of harmonics is slightly different from 
the one used in Hu and White \cite{hw}, and Kodama and Sasaki \cite{KS}. The reason for this
is that in \cite{hw} and \cite{KS} one assumes that all variables represent 
square roots of power spectra. Isotropy can then be assumed.
This procedure will indeed lead to the right ensemble average 
$P(k)$ and $C_\ell$ if the source is coherent (or a decomposition
into coherent eigenmodes has been performed.) However it is not
valid for each realization of an incoherent source.

\section{Tool 2: Unequal-time  correlators}\label{corrs}

The unequal time correlators are defined as  
\begin{equation} 
{\langle \Theta_{\mu\nu}({\bf k},\eta)\Theta^\star_{\alpha\beta} 
({\bf k},\eta ')\rangle}\equiv{\cal C}_{\mu\nu,\alpha\beta}(k,\eta,\eta ') 
\end{equation} 
where $\Theta_{\mu\nu}$ is the stress energy tensor, 
${\bf k}$ is the wave-vector, and $\eta$ and $\eta '$  
are any two (conformal) times. The  
UETCs determine all other 2 point functions, most notably 
CMB and LSS power spectra $C_\ell$ and $P(k)$.  

All correlators between modes at $({\bf k},\eta)$ and
$({\bf k}',\eta')$ will be proportional to $\delta({\bf k}-{\bf k}')$
due to translational invariance. We shall drop this factor in all formulae.
The correlators can also be functions of $k$ alone,
due to isotropy.  Since conjugation corresponds to ${\bf k}\rightarrow -{\bf k}$, isotropy implies that the correlators must be real. Because of incoherence 
the correlators will be generic functions of $\eta$ and $\eta'$. 

Furthermore, the form of the S+V+T decomposition fixes
further the form of the correlators. One can always write down the most
general form of a correlator, and then contract the result with
${\hat k_i}$ or $\delta_{ij}$, wherever appropriate, 
to obtain further conditions. For instance for
any scalar $S$ and vector $V_i$ we could write
\begin{equation}
{\langle S V_i\rangle}=\alpha(k){\hat k_i}
\end{equation}
By contracting with ${\hat k_i}$ we would then conclude that
$\alpha(k)=0$. Proceeding in this way we can show that cross
correlators involving components of different type (S, V, or T)
must be zero. Furthermore for any two vectors $V_i$, $W_i$,
one has
\begin{equation}
{\langle V_i W^*_j\rangle}=
f^{VW}(k,\eta,\eta')(\delta_{ij}- {\hat k_i}{\hat k_j})
\end{equation}
and so instead of 6 functions we have only one. In order to agree
with conventions used in the literature we shall instead define this single
function by means of 
\begin{equation}
f^{VW}(k,\eta,\eta')=\sum_{i}{\langle V_i W^*_i\rangle}
\end{equation}

For correlators involving two tensor quantities one can use similar
arguments to prove the general form 
\begin{eqnarray}
{\langle \Pi^T_{ij} \Pi^{T\star}_{kl}\rangle}&=&
f^{\Pi^T\Pi^T}(k,\eta,\eta')
( \delta_{ij}\delta_{kl} -(\delta_{ik}\delta_{jl}+\delta_{il}\delta_{jk})\nonumber\\
&&-(\delta_{ij}{\hat k_k}{\hat k_l}
+\delta_{kl}{\hat k_i}{\hat k_j}) 
+(\delta_{ik}{\hat k_j}{\hat k_l}
+\delta_{il}{\hat k_j}{\hat k_k}+\nonumber\\
&&\delta_{jk}{\hat k_i}{\hat k_l}+
\delta_{jl}{\hat k_i}{\hat k_k})-{\hat k_i}{\hat k_j}
{\hat k_k}{\hat k_l})
\end{eqnarray}
and again we have a single function rather than 21. Again in order to
comply with conventions in the literature we shall define this single function
by means of 
\begin{equation}
f^{\Pi^T\Pi^T}(k,\eta,\eta')=\sum_{ij}{\langle \Pi^T_{ij} \Pi^{T\star}_{ij}
\rangle}
\end{equation}

Hence there should be 10 scalar correlator functions
\begin{equation}
\begin{array} {cccc} 
f^{\rho^d\rho^d}& f^{\rho^d v^d}& f^{\rho^d p^d}& f^{\rho^d \Pi^S}\\
\cdots&f^{v^d v^d}&f^{v^dp^d} &f^{v^d \Pi^S}\\ 
\cdots&\cdots &f^{p^d p^d}&f^{p^d \Pi^S}\\
\cdots&\cdots &\cdots &f^{\Pi^S\Pi^S}\\
\end{array}
\end{equation}
3 vector correlators:
\begin{equation}
\begin{array} {cc}
f^{\omega^d\omega^d}&f^{\omega^d\Pi^V}\\
\cdots& f^{\Pi^V\Pi^V}
\end{array}
\end{equation}
and a single tensor correlator function $f^{\Pi^T\Pi^T}$.

In general these functions are functions of $(k,\eta,\eta')$,
and this is indeed the case during the matter radiation transition.
However well into the matter and radiation epochs there is scaling,
and these functions may be written as:
\begin{equation}
f^{XY}(k,\eta,\eta')={F^{XY}(x,x')\over \sqrt{\eta\eta '}}
\end{equation}
where $XY$ represents any pair of superscripts considered above,
and $x=k\eta$ and $x'=k\eta'$. The above scaling form results
from the dimensional analysis argument: $[\Theta_{\mu\nu}(x)]=1/L^2$, 
$[\Theta_{\mu\nu}(k)]=L$, $[\delta({\bf k})]=L^3$, $[f^{XY}]=1/L$.
We have used units where $G=c=1$.

\section{Causality constraints}\label{caus}

From analyticity conditions \cite{tps} we can use isotropy and symmetry
\cite{tps,uzan} to derive general expansions in $\bf k$ around
$k_i=0$.  This is usually very helpful to check numerical results. 
We find:

\bea
\vev{\Theta_{00}(k,\eta)\Theta_{00}(k,\eta')}&=&X\nonumber\\
\vev{\Theta_{00}(k,\eta)\Theta_{0i}(k,\eta')}&=&ik_iY\nonumber\\
\vev{\Theta_{00}(k,\eta)\Theta_{ij}(k,\eta')}&=&V\delta_{ij}+Wk_ik_j\nonumber\\
\vev{\Theta_{0i}(k,\eta)\Theta_{0j}(k,\eta')}&=&T\delta_{ij}+Uk_ik_j\nonumber\\
\vev{\Theta_{0i}(k,\eta)\Theta_{jk}(k,\eta')}&=&i [ Qk_i\delta_{jk} +
R(k_j\delta{ik}+k_k\delta_{ij})\nonumber\\
&&+Sk_ik_jk_k ]\nonumber\\
\vev{\Theta_{ij}(k,\eta)\Theta_{kl}(k,\eta')}&=&A\delta_{ij}\delta_{kl}
+ B(\delta_{ik}\delta_{jl}+\delta{il}\delta{jk})\nonumber\\
&&+C(k_ik_j\delta{kl}+k_kk_l\delta{ij})\nonumber\\
&&+D(k_ik_k\delta{jl}+k_ik_l\delta{jk}
+k_jk_l\delta{ik}\nonumber\\
&&+k_jk_k\delta{il}) + Ek_ik_jk_kk_l
\eea
where A, B etc. are functions independent of $k$.
By applying the scalar, vector and tensor decomposition we can then find
the specific form of the correlators in our formalism to obtain relations
between the correlators
\bea
f^{\rho\rho}&=&X\nonumber\\
f^{\rho p}&=&V+{1\over 3}k^2W\nonumber\\
f^{\rho v}&=&kY\nonumber\\
f^{\rho\Pi^S}&=&{3\over 2}W(k^2-{1\over 3})\nonumber\\
f^{pp}&=&A+{2\over3}B+{2\over 3}k^2(C+{2\over 3}D)+{1\over 9}k^4E\nonumber\\
f^{pv}&=&k(Q+{2\over 3}R) +{1\over 3}Sk^2\nonumber\\
f^{p\Pi^S}&=&(C-{4\over3}D)k^2+{1\over 3}Ek^4\nonumber\\
f^{vv}&=&-(T+k^2U)\nonumber\\
f^{v\Pi^S}&=&k(2R-Sk^2)\nonumber\\
f^{\Pi^S\Pi^S}&=&3B+4Dk^2+Ek^4\nonumber\\
f^{\omega\omega}&=&2T\nonumber\\
f^{\omega\Pi^S}&=&2Rk\nonumber\\ 
f^{\Pi^V\Pi^V}&=&2B\nonumber\\ 
f^{\Pi^T\Pi^T}&=&4B
\eea
We can now derive ${\bf{k}}\rightarrow 0$ constraints on the correlators.
These may 
help to complete the functions $F^{XY}$ near the origin, where they may
not be accessible from simulations due to low sampling of
points on the lattice. Specifically, we get ratios between the scalar,
vector and tensor anisotropic stresses and between the momentum and
vorticity,
\bea\label{stress}
f^{\Pi^S\Pi^S}:f^{\Pi^V\Pi^V}:f^{\Pi^T\Pi^T}
&=&3:2:4\nonumber\\
f^{vv}:f{\omega\omega}&=&1:2
\eea
to zeroth order and,
\bea
f^{v\Pi^S}&\approx&f^{\omega_i\Pi^V_i}
\eea
to first order.

\section{Tool 3: Simulation determination of the correlators}\label{sim}
\begin{figure}[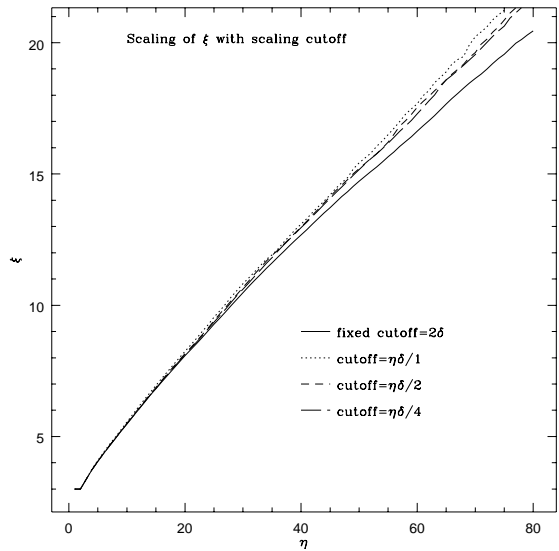]
\begin{center}
\leavevmode \epsfysize=8cm  \epsfbox{plot.ps}
\end{center}
\caption{Scaling of the correlation length $\xi$ for various loop
cut off sizes.}
\label{fig1}
\end{figure}

Having developped the necessary analytical tools we now describe
how to apply them to simulations. We consider the work in \cite{chm}.
The simulation starts with a network of strings obtained using the algorithm
developed in \cite{VV1} which simulates the symmetry breaking of the 
underlying field by selecting random phases for the field at each point in
the lattice. Strings are identified on points in the lattice where the phase
of the field has a non-zero winding number. The strings are described by
their position $\bX(\sigma,\eta)$ where $\sigma$ is a parameter running
along the string and $\eta$ is conformal time. By imposing the following
gauge conditions 
\begin{equation}
\bXd.\bXp=0 \qquad \bXd^2+\bXp^2=1
\end{equation}
where dots and primes denote derivatives with respect to $\eta$ and $\sigma$, respectively,
the equation of motion takes on the simple form of the wave equation  
\begin{equation}
\dprime \bX^2 +\ddot{\bX}^2 = 0
\end{equation}
 and can be discretised on the lattice so that all the possible velocities
take integer values only\cite{fst,mairi}. This greatly increases the
accuracy and speed of 
string network codes based on this algorithm. 
The network is then evolved using the discretised equations of motion and
intercommuting relations. To simulate the extraction of energy from the
system due to the decay of the string loops into gravitational radiation
and/or particles, loops of a minimum size are excised from the
simulation at each time step. This also ensures that the network scales with
respect to the conformal time $\eta$ which enables us to extend the
dynamical range of the resulting correlation functions beyond the limited
range covered in the simulation. Fig.~\ref{fig1} shows how the correlation
length defined by 
$\xi={\sqrt{\mu/\rho_l}}$
scales as a function of conformal time. Here, $\rho_l$ is the density of 'long' strings.

We performed simulations
with box sizes ranging from $128^3$ to $600^3$, with a cut-off on the loop
size of two links. Realisation averages were carried out with $256^3$ boxes
once it was determined that the general form of the correlators scaled very
accurately with box size. To evaluate the UETCs from the simulations we
selected times in the range $0.1 N<t<N/4$, where $N$ is the box size,
when we were sure that
the string network was scaling, and when boundary effects are 
still excluded by causality. At each of the time steps the vectors $\bXd$ and
$\bXp$
from each point along the strings in the network were stored. From
this we obtained the time evolution of the network's stress-energy
tensor$\Theta_{\mu\nu}$ at each point on the lattice using  
\begin{equation}
\Theta_{\mu\nu}({\bf x})={\mu\int d\si
(\Xd^{\mu}\Xd^{\nu}-\Xp^{\mu}\Xp^{\nu})\delta^3({\bf x}-\bX(\si,\eta))} 
\end{equation}

The Fast Fourier Transforms of all 10 independent components of the string 
stress-energy were then decomposed into irreducible scalar, vector and
tensor modes using eqns in Section (\ref{svt}). 
This resulted in all the SVT components being
computed directly from the simulation without making assumptions on energy
conservation and on the details of energy dissipation from the string
network. The drawback of obtaining all the components directly in such a
manner is that the process becomes computationally intensive for even
modestly sized simulations, e.g. $256^3$, as, in effect, one has to deal with
$\approx 10$ times the number of variables at each point on the lattice.  

By cross-correlating the decomposed stress-energy components from a central
time with those from all the stored time steps the 14 independent UETCs
$f^{XY}(k,\eta,\eta')$ were computed.

\begin{figure}[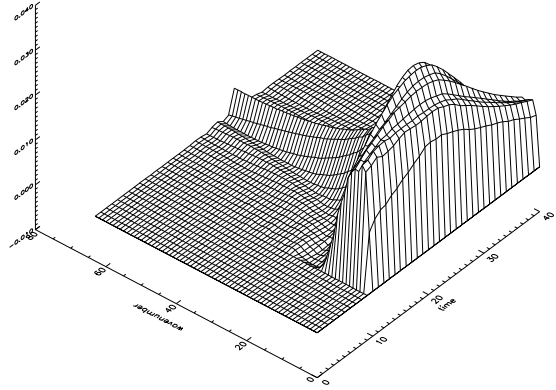]
\begin{center}
\leavevmode \epsfysize=8cm  \epsfbox{pow.ps}
\end{center}
\caption{The function ${\langle |\rho^{d}|^2\rangle}(k,25,\eta)$.}
\label{first}
\end{figure}

\begin{figure}[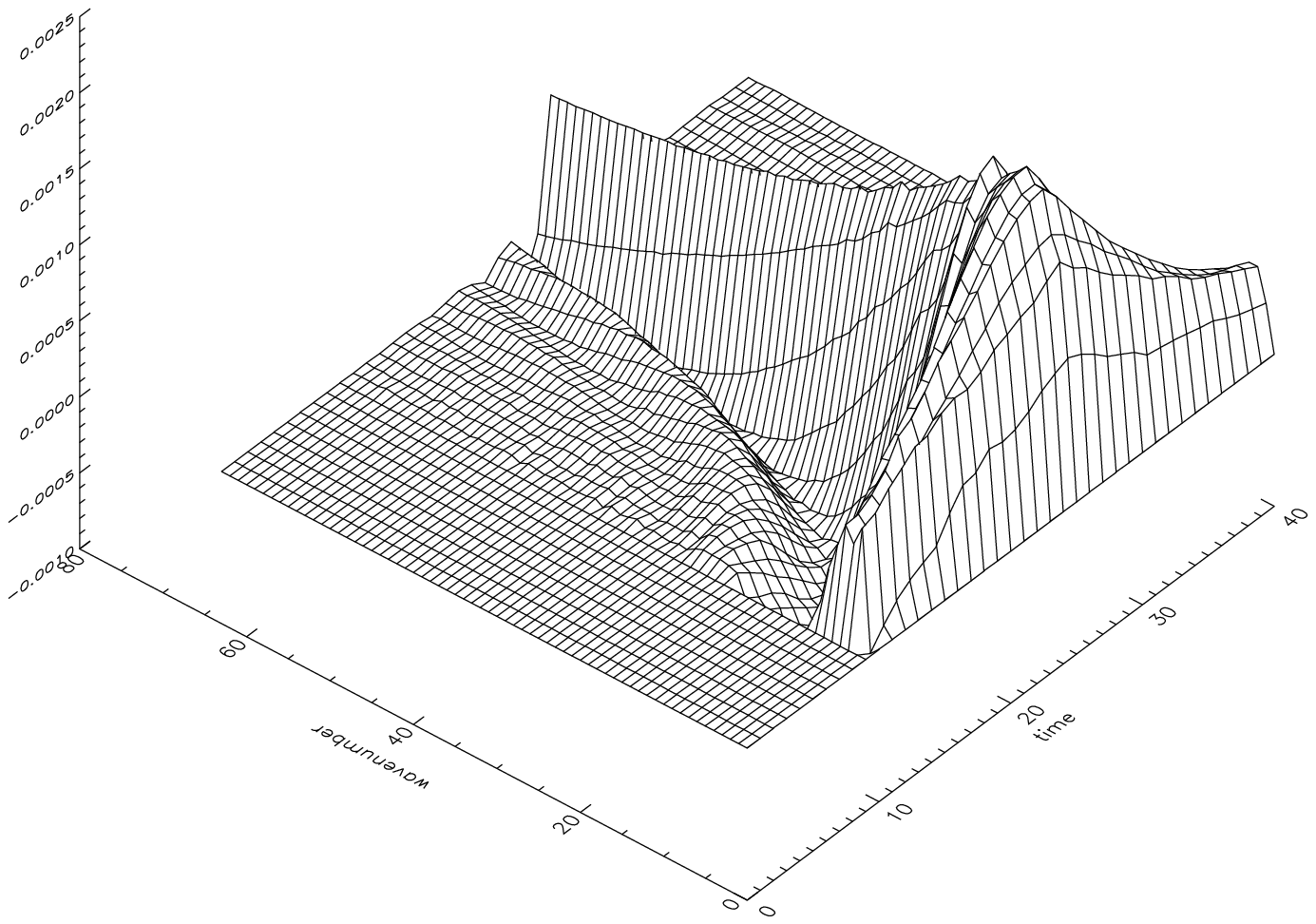]
\begin{center}
\leavevmode \epsfysize=8cm  \epsfbox{U-U.ps}
\end{center}
\caption{The function ${\langle |v^d|^2 \rangle}(k,25,\eta)$.}
\label{fig2}
\end{figure}

\begin{figure}[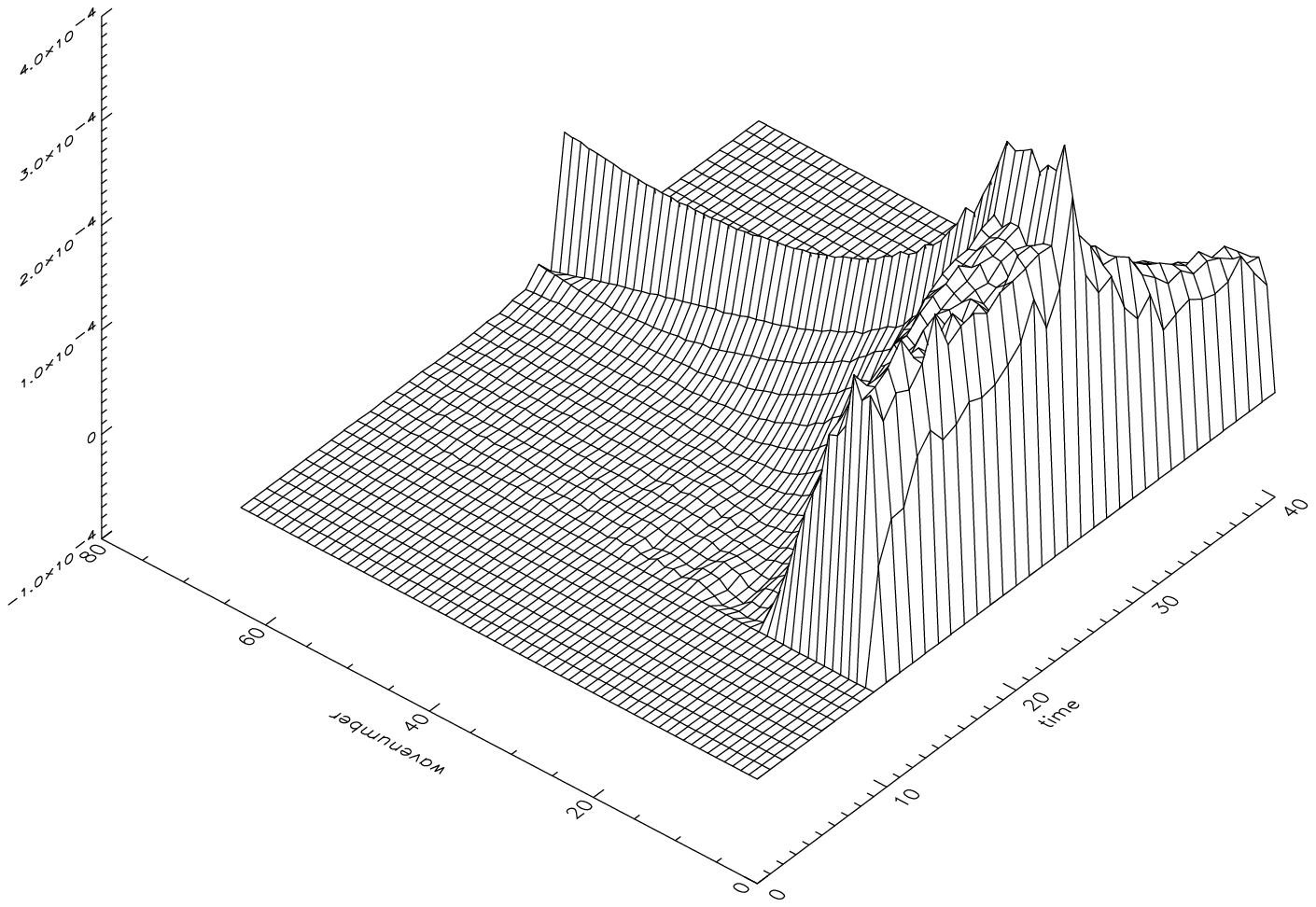]
\begin{center}
\leavevmode \epsfysize=8cm  \epsfbox{p-p.ps}
\end{center}
\caption{The function ${\langle |p^d|^2\rangle}(k,25,\eta)$.}
\label{fig3}
\end{figure}

\begin{figure}[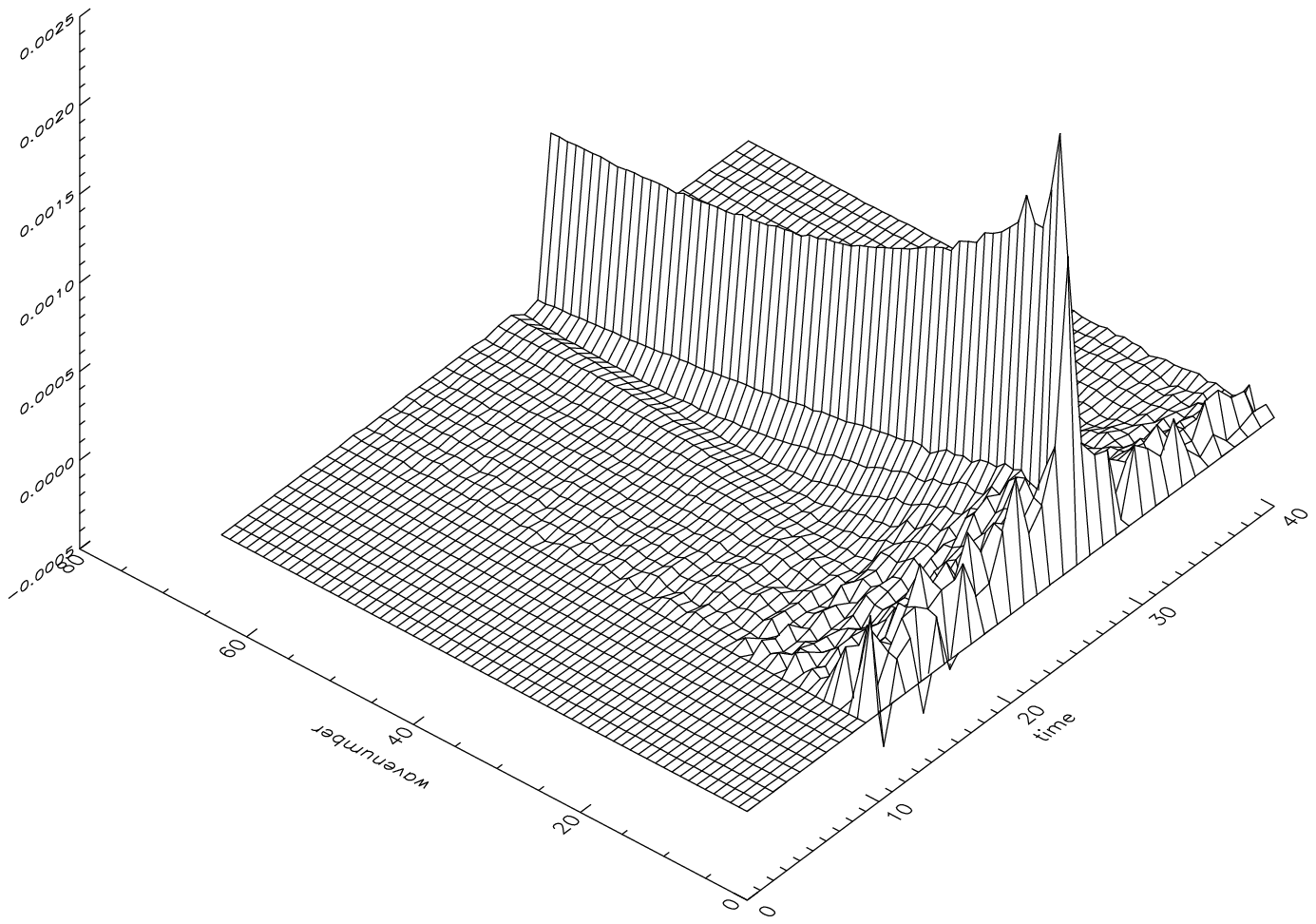]
\begin{center}
\leavevmode \epsfysize=8cm  \epsfbox{sca.ps}
\end{center}
\caption{The function ${\langle |\Pi^S|^2\rangle}(k,25,\eta)$.}
\label{fig4}
\end{figure}

\begin{figure}[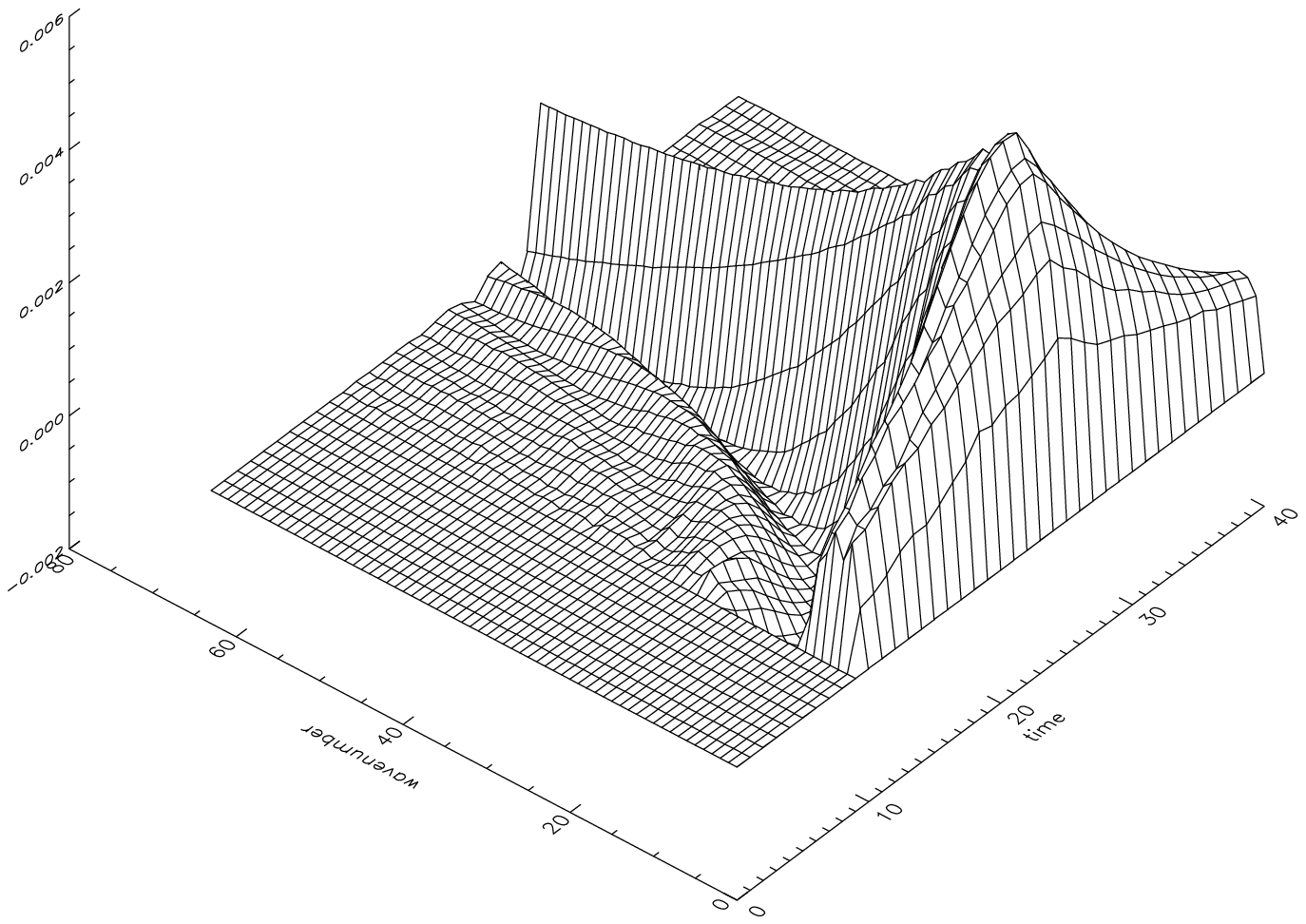]
\begin{center}
\leavevmode \epsfysize=8cm  \epsfbox{V.ps}
\end{center}
\caption{The function $\sum_i{\langle |\omega^d_i|^2 \rangle}(k,25,\eta)$.}
\label{fig5}
\end{figure}

\begin{figure}[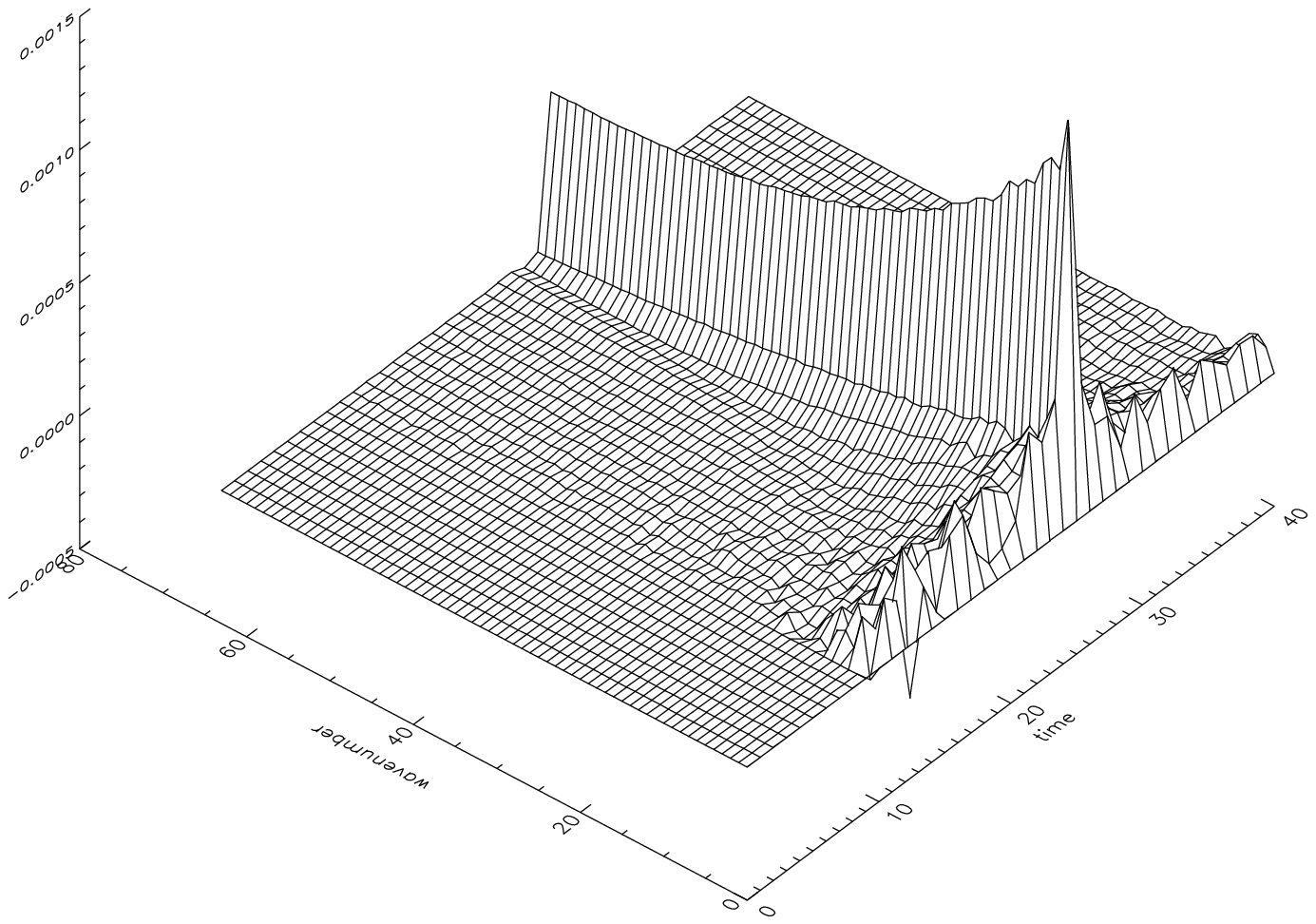]
\begin{center}
\leavevmode \epsfysize=8cm  \epsfbox{vec.ps}
\end{center}
\caption{The function $\sum_i{\langle |\Pi^V_i|^2\rangle}(k,25,\eta)$.}
\label{fig6}
\end{figure}

\begin{figure}[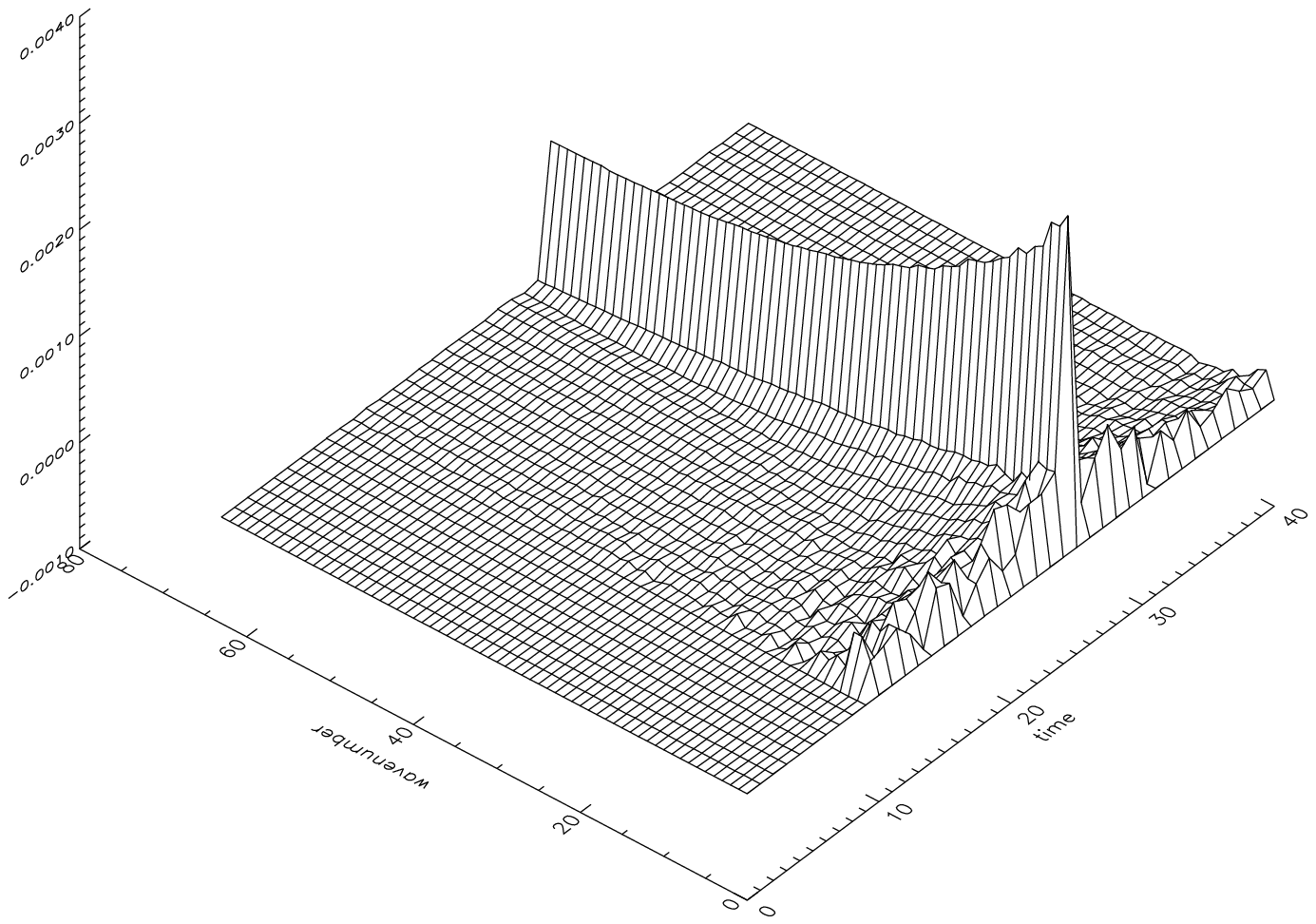]
\begin{center}
\leavevmode \epsfysize=8cm  \epsfbox{ten.ps}
\end{center}
\caption{The function $\sum_{ij}{\langle |\Pi^T_{ij}|^2\rangle}(k,25,\eta)$.}
\label{fig7}
\end{figure}

\begin{figure}[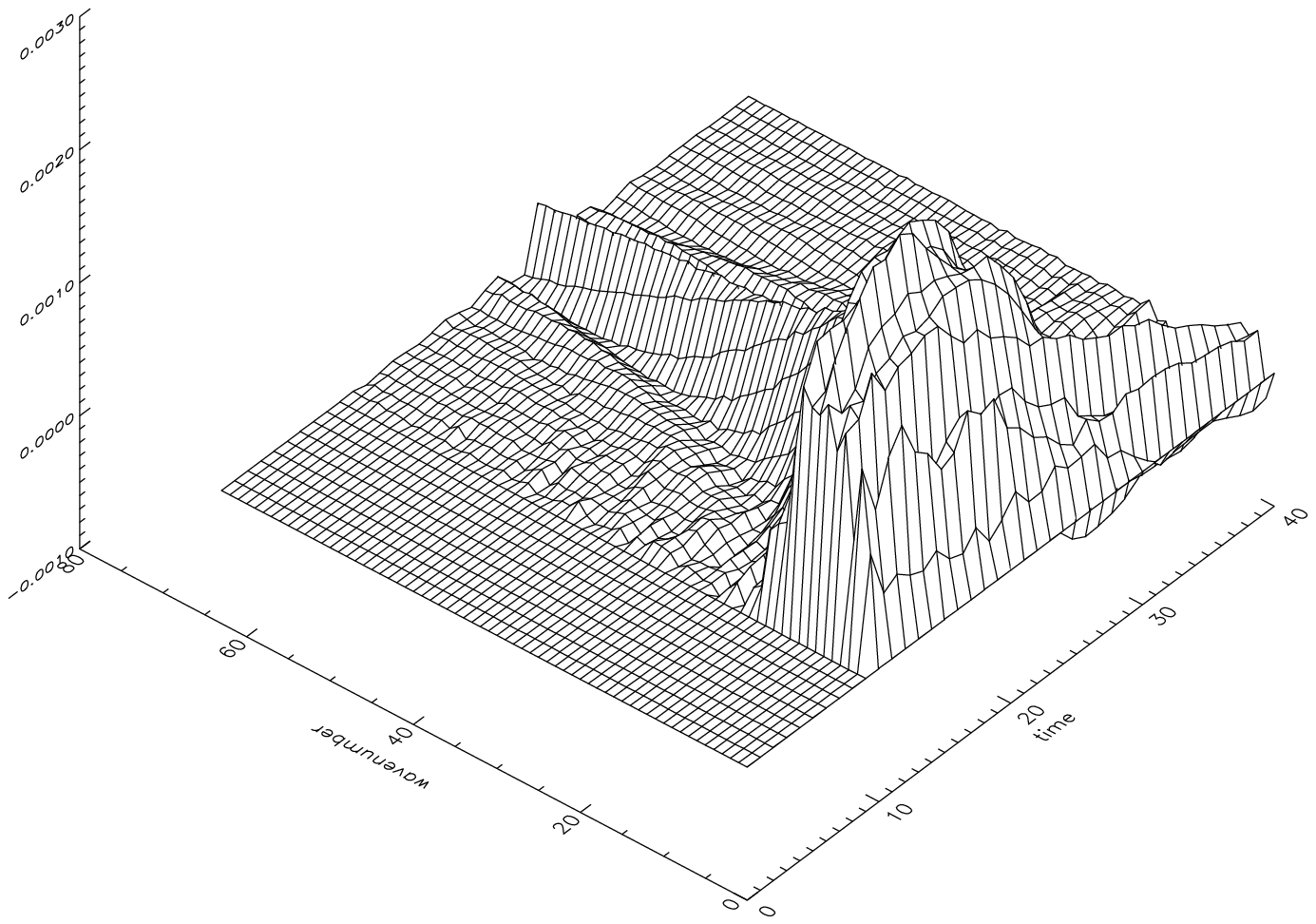]
\begin{center}
\leavevmode \epsfysize=8cm  \epsfbox{rho-U.ps}
\end{center}
\caption{The function ${\langle \rho^d v^{d\star}\rangle}(k,25,\eta)$.}
\label{fig8}
\end{figure}

\begin{figure}[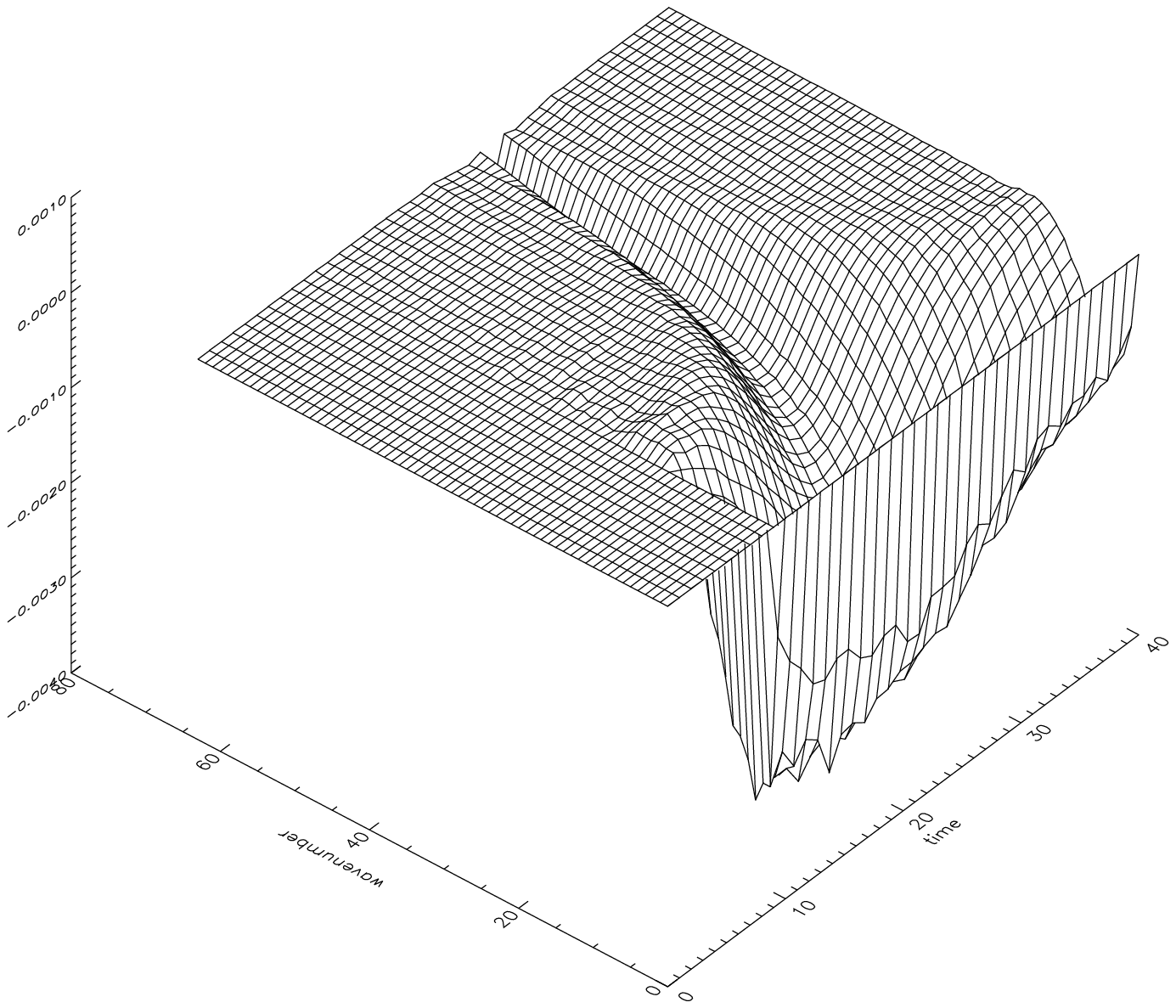]
\begin{center}
\leavevmode \epsfysize=8cm  \epsfbox{rho-p.ps}
\end{center}
\caption{The function ${\langle \rho^d p^{d\star} \rangle}(k,25,\eta)$.}
\label{fig9}
\end{figure}

\begin{figure}[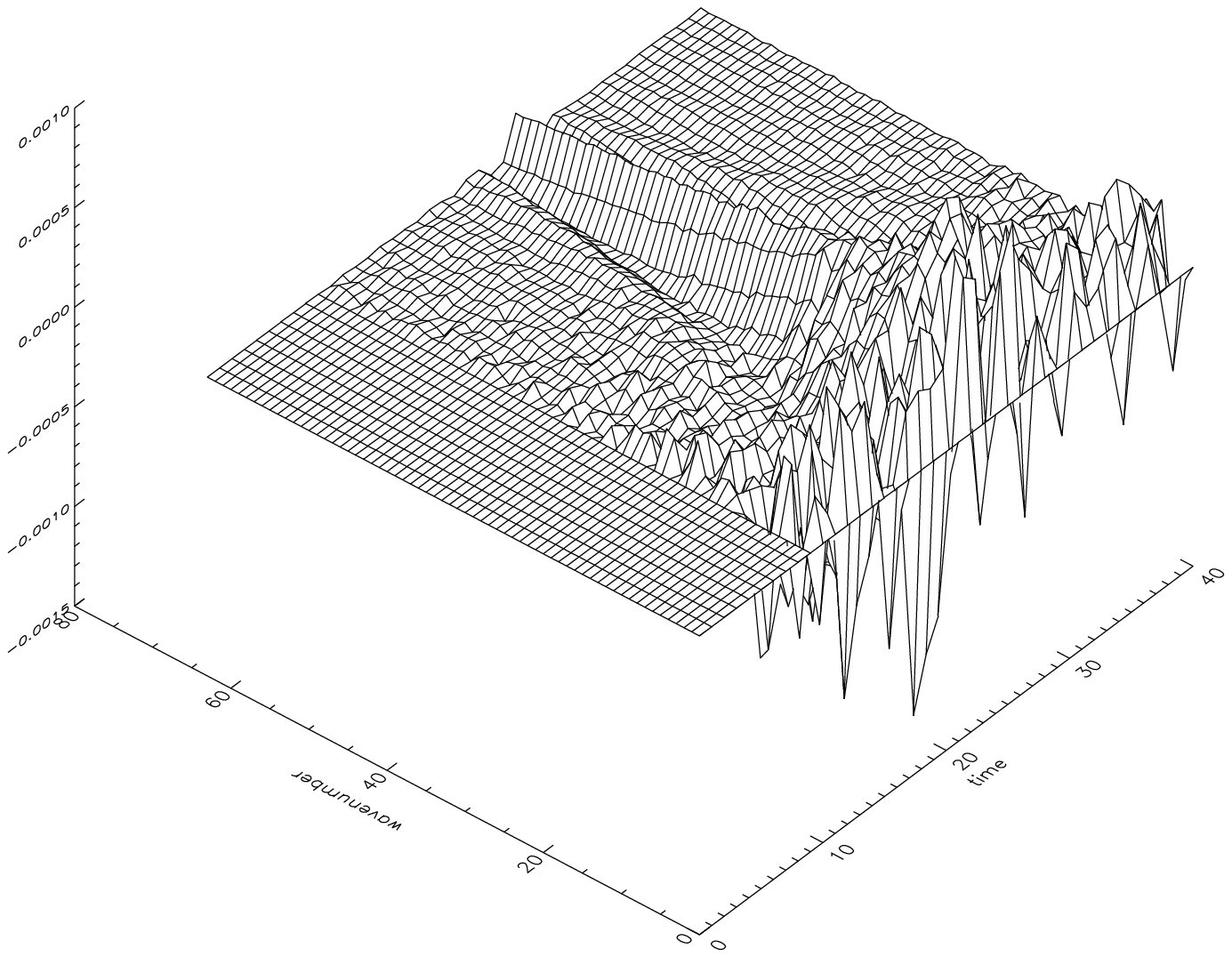]
\begin{center}
\leavevmode \epsfysize=8cm  \epsfbox{rho-sca.ps}
\end{center}
\caption{The function ${\langle \rho^d \Pi^{S\star}\rangle}(k,25,\eta)$.}
\label{fig10}
\end{figure}

\begin{figure}[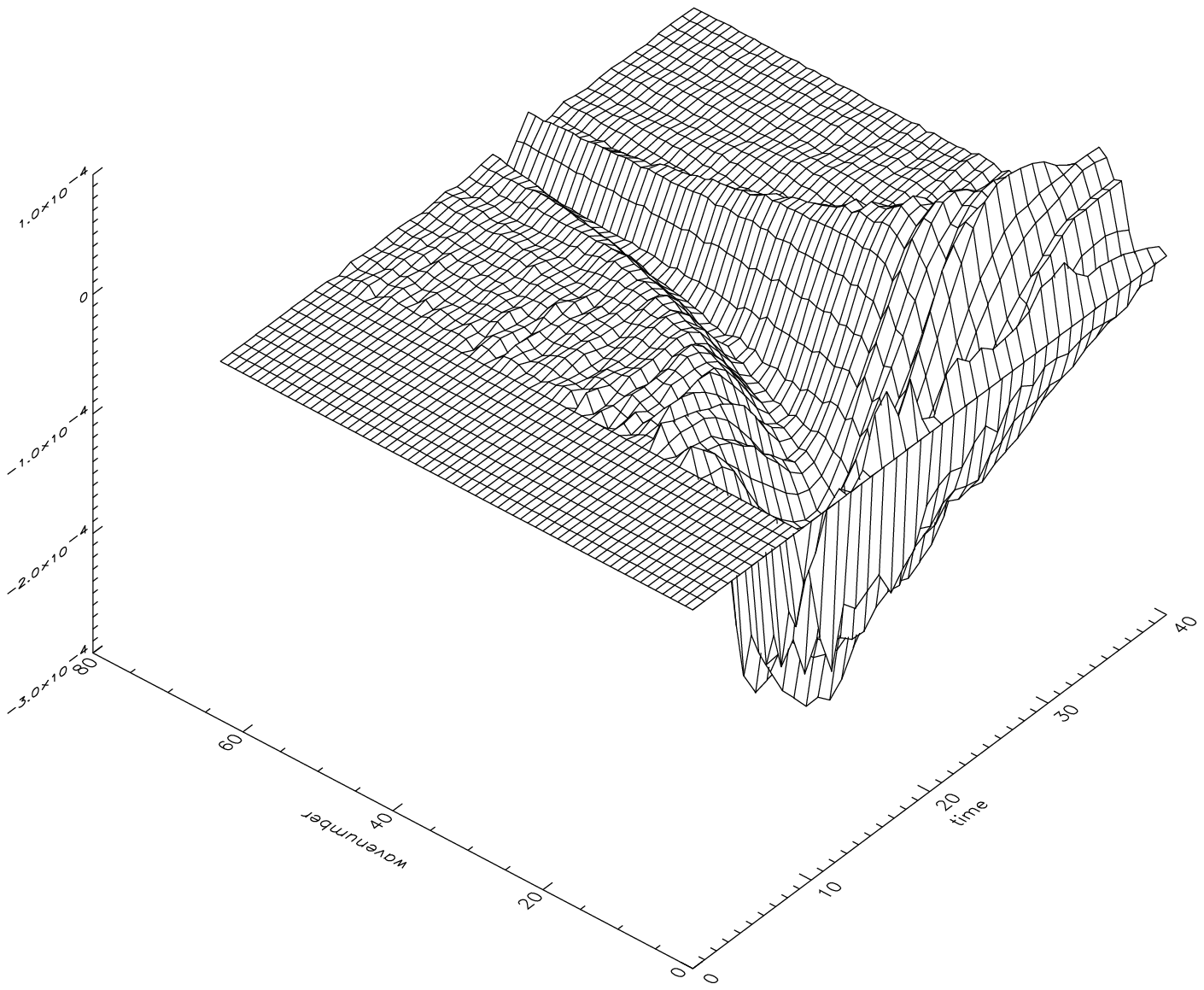]
\begin{center}
\leavevmode \epsfysize=8cm  \epsfbox{p-U.ps}
\end{center}
\caption{The function ${\langle p^d v^{d\star}\rangle}(k,25,\eta)$.}
\label{fig11}
\end{figure}

\begin{figure}[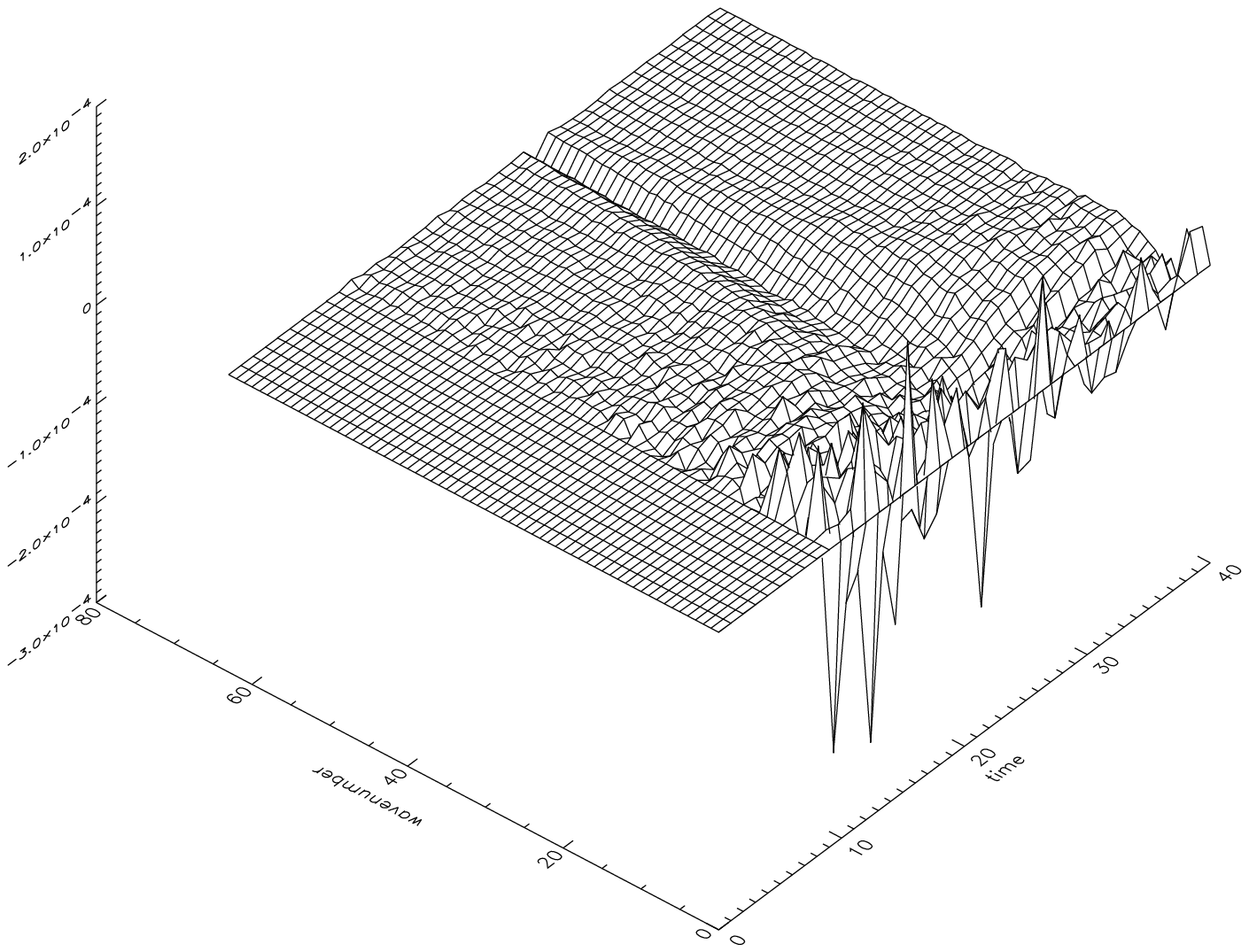]
\begin{center}
\leavevmode \epsfysize=8cm  \epsfbox{p-sca.ps}
\end{center}
\caption{The function ${\langle p^d \Pi^{S\star}\rangle}(k,25,\eta)$.}
\label{fig12}
\end{figure}

\begin{figure}[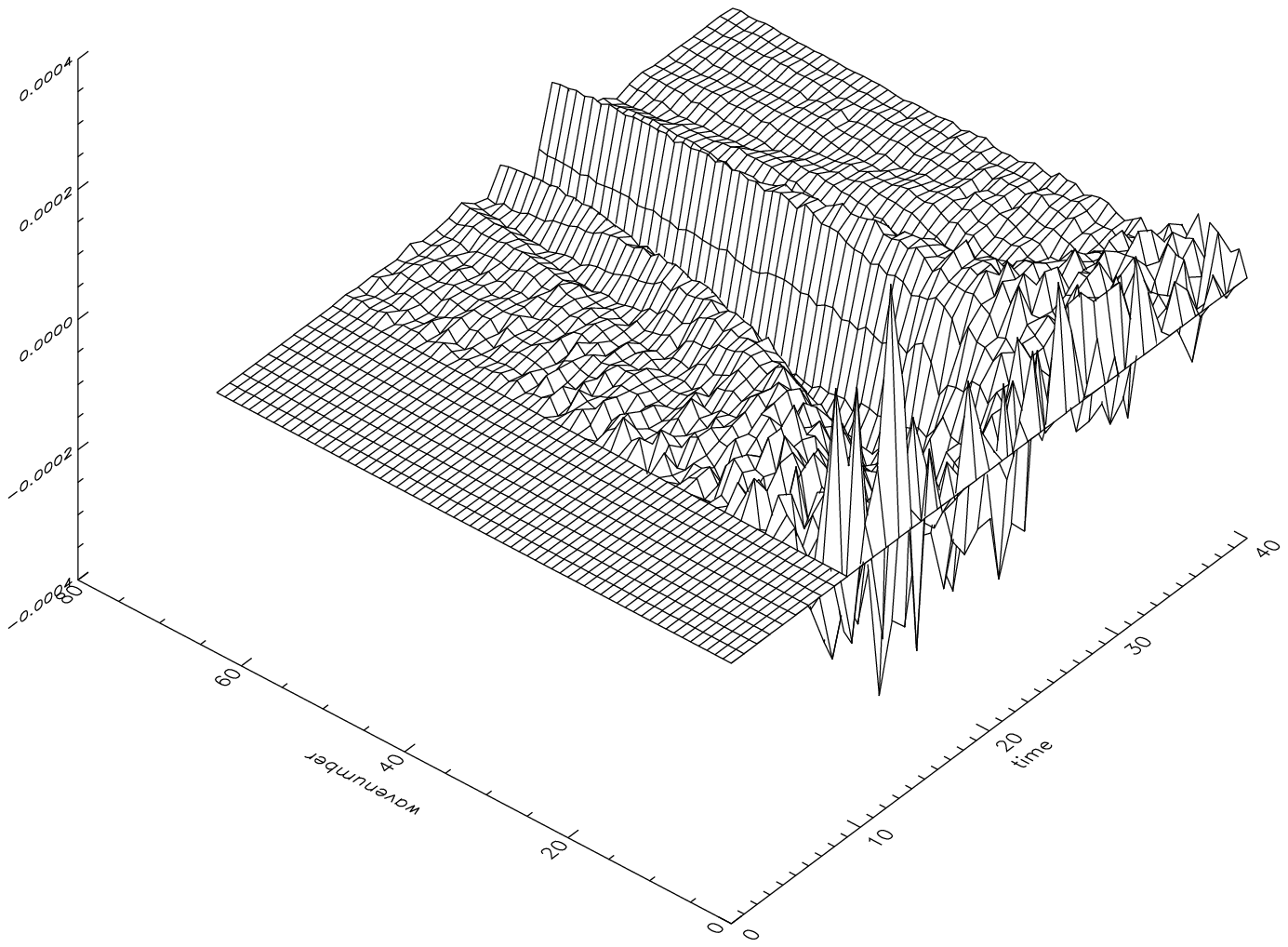]
\begin{center}
\leavevmode \epsfysize=8cm  \epsfbox{U-sca.ps}
\end{center}
\caption{The function ${\langle v^d\Pi^{S\star}\rangle}(k,25,\eta)$.}
\label{fig13}
\end{figure}

\begin{figure}[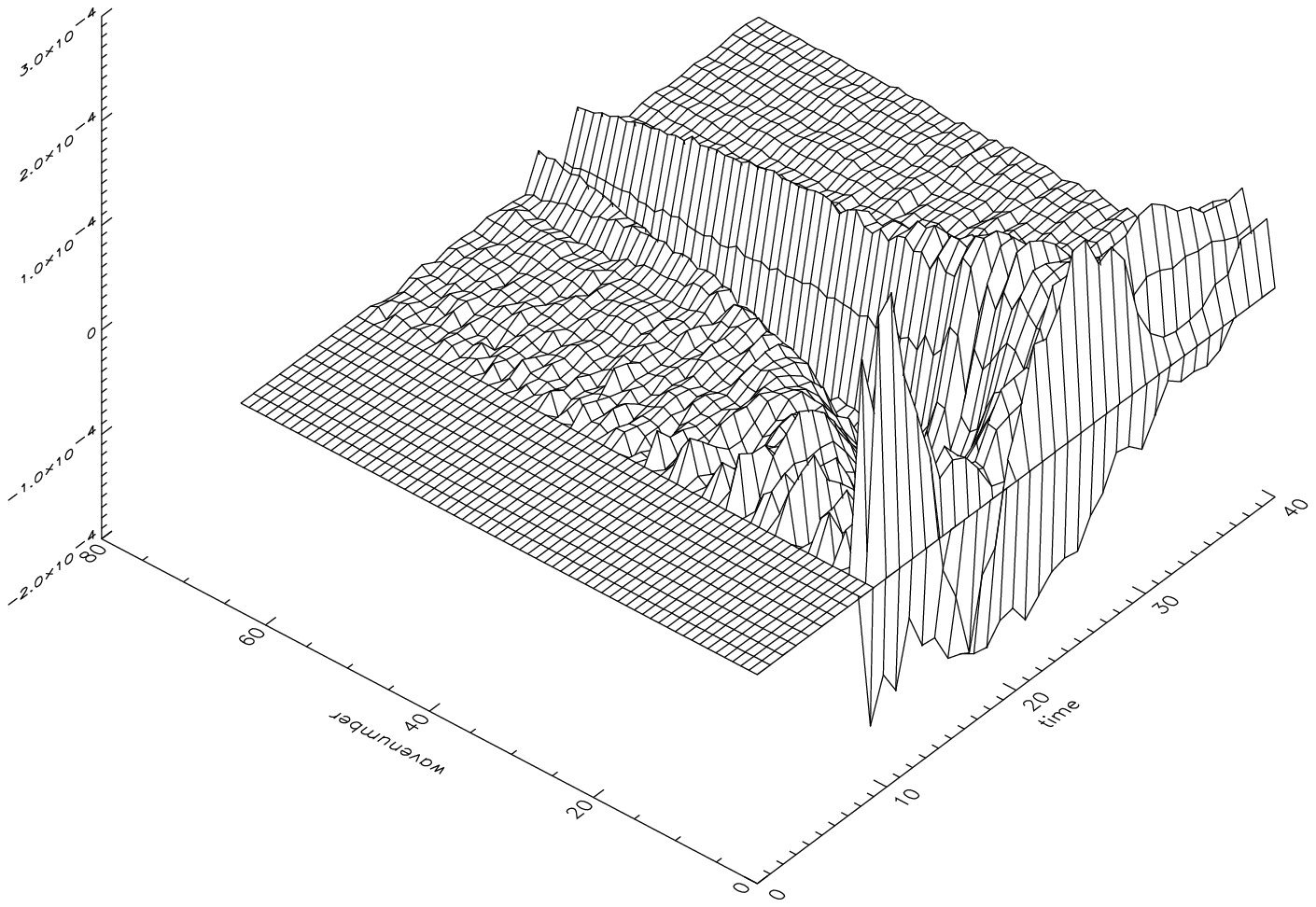]
\begin{center}
\leavevmode \epsfysize=8cm  \epsfbox{v-V.ps}
\end{center}
\caption{The function $\sum_i{\langle \omega^d_i \Pi^{V\star}_i\rangle}(k,25,\eta)$.}
\label{last}
\end{figure}

In Figs.\ref{first}-\ref{last} we display the forms of the various functions 
$f^{XY}(k,25,\eta)$ where $k$, $\eta$ and $\eta'$ are in lattice units.
This is enough to infer the scaling functions $F^{XY}(x,x')$.
One can see that these correlators fall off very quickly away from the
diagonal, a phenomenon known as incoherence \cite{inc} as we explained
above. Incoherence
determines whether or not we have enough dynamical range to compute
the UETCs: if we see the fall off completly clearly we have enough
dynamical range!

\begin{figure}[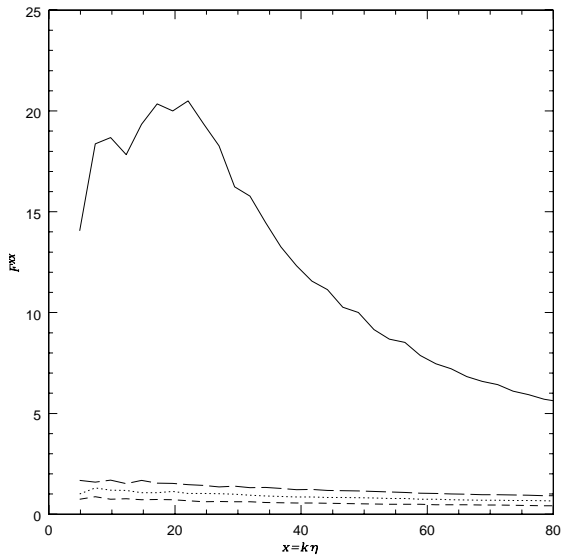]
\begin{center}
\leavevmode \epsfysize=8cm  \epsfbox{rat.ps}
\end{center}
\caption{This plot shows the scaling autocorrelation of the energy
density, the scalar vector and tensor anisotropic stresses.
We see that the energy density dominates all the other components.}
\label{fig15}
\end{figure}

\begin{figure}[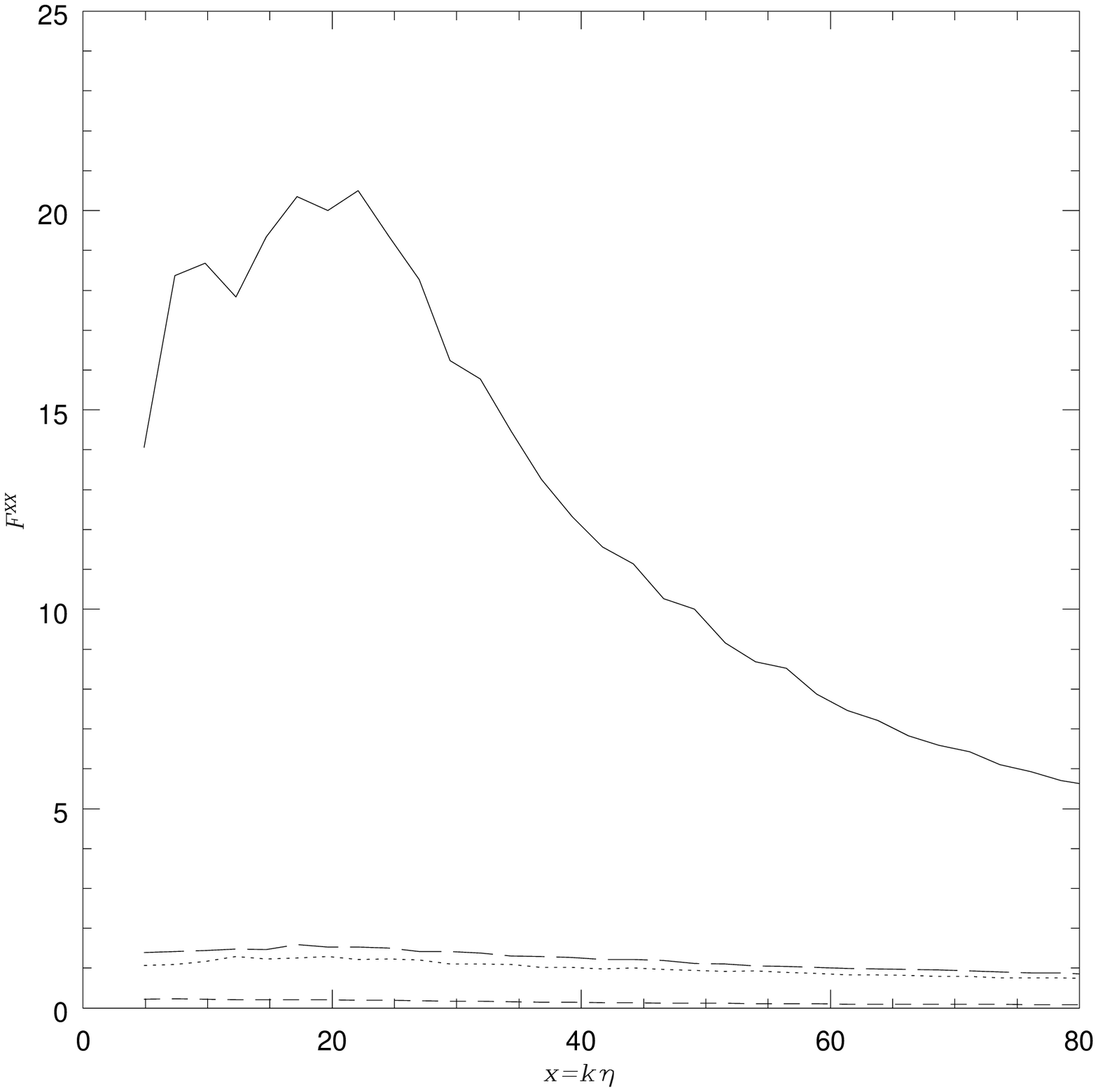]
\begin{center}
\leavevmode \epsfysize=8cm  \epsfbox{vel.ps}
\end{center}
\caption{Same but for the energy density, the pressure, the
velocity and the vorticity}
\label{fig16}
\end{figure}

A striking feature  
of our results is the dominance of  
$\Theta_{00}$ over all other components (see Figs.~\ref{fig15}
and \ref{fig16}).
 The string anisotropic  
stresses are in the predicted \cite{tps} ratios  
$|\Theta^S|^2:|\Theta^V|^2:|\Theta^T|^2$ of $3:2:4$, 
as $k\tau\rightarrow 0$.  
However $|\Theta_{00}|^2\gg | \Theta^S|^2$, and so scalars dominate 
over vectors and tensors. Also the energy density power spectrum 
rises from a white noise tail at $k\tau\approx 0$ into a peak 
at $k\tau\approx 20$, after which it falls off. Sub-horizon 
modes are therefore of great importance. 
These features consistently appeared for all  
box sizes, and are independent of the cutoff size imposed 
on the loops.

\section{Tool 4: Decomposition into eigenmodes}\label{eig}
The UETCs $c_{\mu\nu,\alpha\beta}(k\tau,k\tau ')$ may be diagonalised  
\cite{pst} and written as 
\begin{equation}\label{eig1} 
c_{\mu\nu,\alpha\beta}(k\tau,k\tau ')={\sum_i}\lambda^{(i)} 
v^{(i)}_{\mu\nu}(k\tau)v^{(i)}_{\alpha\beta}(k\tau') 
\end{equation} 
where $\lambda^{(i)}$ are eigenvalues. In general, defects are  
incoherent sources for perturbations \cite{inc}, which means that  
this matrix does not 
factorize into the product of two vectors  
$v_{\mu\nu}(k\tau)v_{\alpha\beta}(k\tau')$. Standard codes solving for   
CMB and LSS power spectra assume coherence.  However  
we see that an incoherent source may be represented as an 
incoherent sum of coherent sources. We may therefore  
feed each eigenmode into standard codes \cite{cmbfast} to find the  
$C^{(i)}_\ell$ and $P^{(i)}(k)$ associated with each mode. 
The series $\sum \lambda^{(i)} 
C_\ell^{(i)}$ and $\sum \lambda^{(i)}P^{(i)}(k)$ provide  
convergent approximations to the power spectra.

\begin{figure}[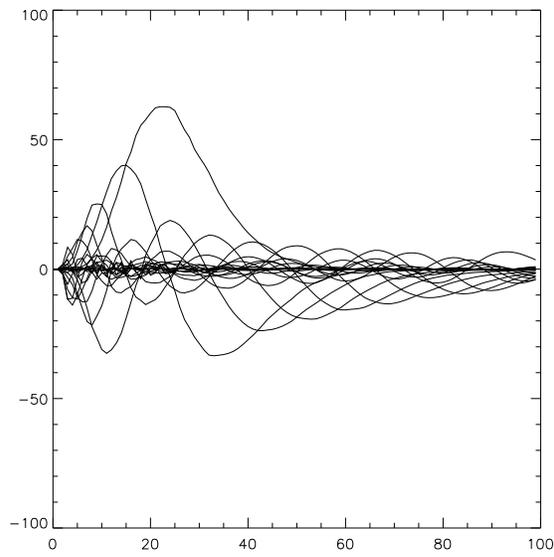]
\begin{center}
\leavevmode \epsfysize=8cm  \epsfbox{eigenpow.ps}
\end{center}
\caption{Decomposition into eigenmodes of the energy density.}
\label{fig17}
\end{figure}

\begin{figure}[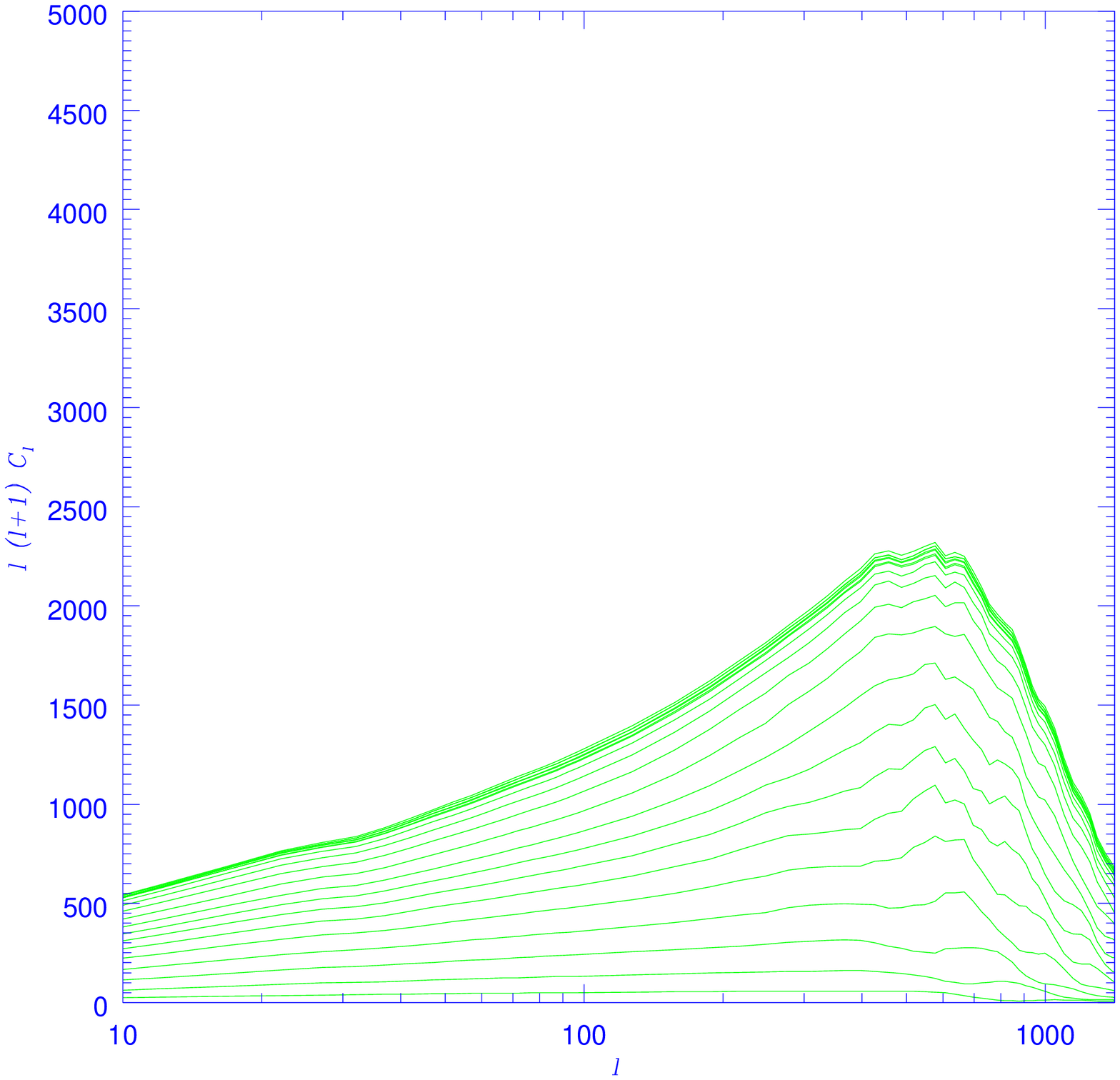]
\begin{center}
\leavevmode \epsfysize=8cm  \epsfbox{cl.ps}
\end{center}
\caption{The $C_\ell$ spectrum and inferred from 20 eigenmodes.}
\label{fig18}
\end{figure}

In Fig.~\ref{fig17} we show the eigenmodes corresponding to the energy
density $\Theta_{00}$. We see that the leading eigenmode carries
the mark of the peak in the energy power spectrum at $x \approx 20$.
The other eigenmodes become very small very quickly and change sign
with higher and higher frequency. If we compute the $C_\ell$ power
spectrum for each of these modes, and sum the series, we find
quick convergence shortly after 20 modes have been included
(see Fig.~\ref{fig18}).

\section{Some results}\label{result}

The methods described above supplied a variety of interesting results
for cosmic strings. They show that local strings, unlike global defects
have a Doppler peak. The main uncertainty seems to be in the details
of how strings loose energy (see e.g. \cite{vinc,vhs}). String decay products are clearly the most uncertain aspect  
of cosmic string theory. By measuring the full 14 UETCs associated 
with long strings, we assume nothing 
about decay products when extracting information from simulations 
The simulations will then also place constraints 
upon the decay products. 

\begin{figure}[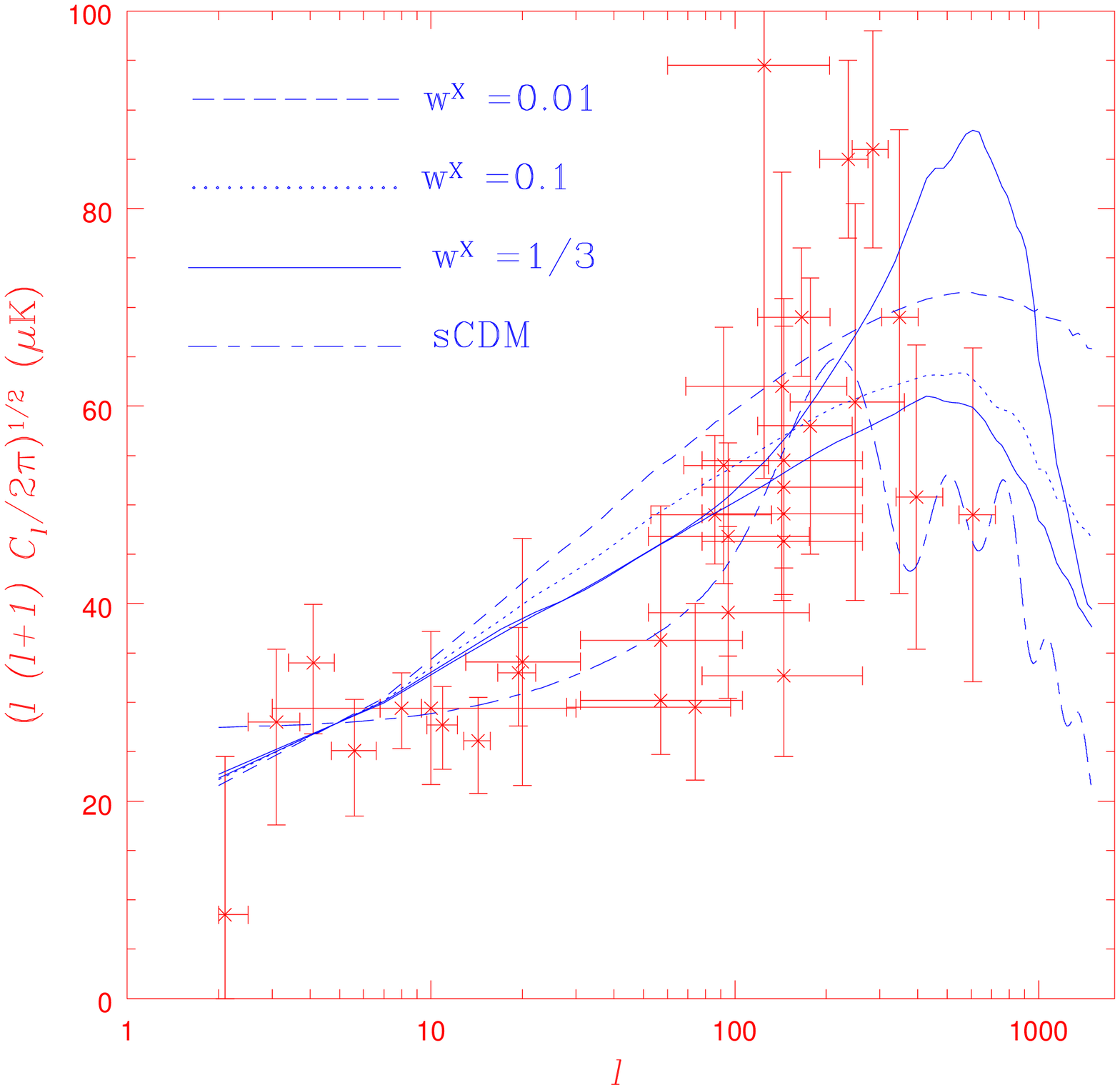] 
\begin{center}
\leavevmode \epsfysize=8cm  \epsfbox{res1.ps}
\end{center}
\caption{The CMB power spectra predicted by cosmic strings decaying 
into loop and radiation fluids with $w^X=1/3, 0.1, 0.01, 0$.  
We have plotted $(\ell(\ell+1)C_\ell/2\pi)^{1/2}$ in $\mu K$, 
and superposed several experimental points. The higher curve  
corresponding to $w^X=1/3$ shows what happens if 5\% of the 
energy goes into the radiation 
fluid.} 
\label{res1} 
\end{figure}

In Fig.~\ref{res1} we plot $\surd[\ell(\ell+1)C_\ell/2\pi]$,   
setting the Hubble constant to 
$H_0=50$ Km sec$^{-1}$ Mpc$^{-1}$, the baryon fraction to 
$\Omega_b=0.05$, and  assuming a flat geometry, no cosmological 
constant, 3 massless neutrinos, standard recombination, 
and cold dark matter.  
We also superimpose current experimental points.  
The most interesting feature is  
the presence of a reasonably high Doppler peak at $\ell=400-600$,  
following a pronouncedly tilted large angle plateau
(cf. \cite{per}). 
This feature sets local strings apart from global defects. 
It puts them in a better shape to face the current data. 
 
The CMB power spectrum is relatively insensitive to  
the equation of state of the extra fluid.  
We have plotted results for $w^X=1/3, 0.1, 0.01$.  
Dumping some energy into CDM has negligible effect. 
Small dumps into  
baryon and radiation fluids, on the contrary, 
boost the Doppler peak very strongly. We plotted the effect  
of dumping 5\% of the energy into the radiation fluid. 
 
\begin{figure}[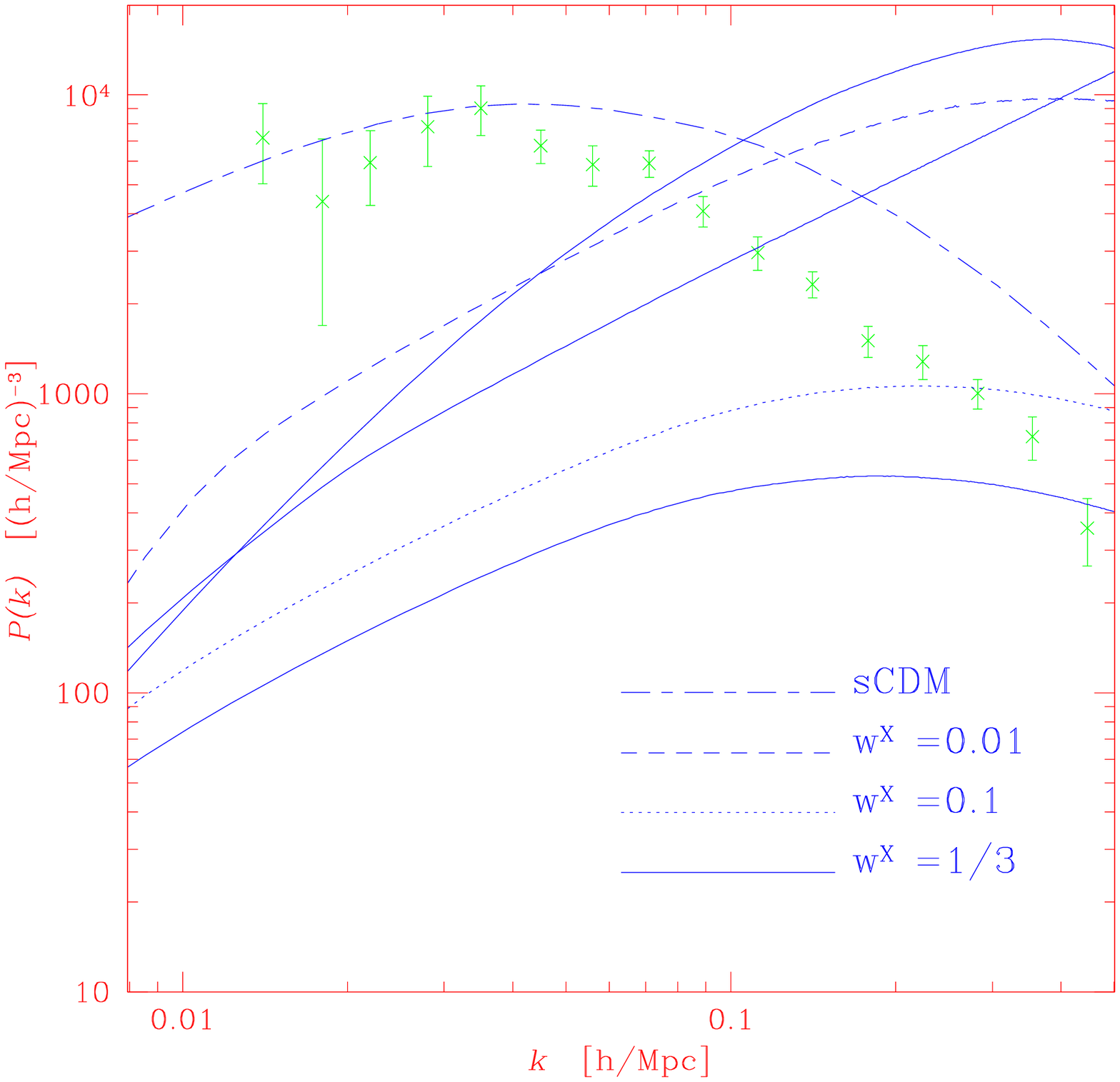] 
\begin{center}
\leavevmode \epsfysize=8cm  \epsfbox{res2.ps}
\end{center}
\caption{The power spectrum in CDM fluctuations for 
cosmic strings, with $w^X=0.01,0.1,1/3$. We plotted 
also the standard CDM scenario prediction and points inferred 
by Peacock and Dodds from galaxy surveys. The top 2 $w^X=1/3$ curves 
correspond to a 5\% transfer into CDM, and a 20\% transfer into  
baryons (top).} 
\label{res2} 
\end{figure}

The LSS power spectra on the other hand is strongly dependent  
on $w^X$. 
In Fig.~\ref{res2} we plotted the CDM power spectrum $P(k)$  
together with experimental points as in \cite{pdodds}. 
The normalization has been fixed by COBE data points. 
We see that the peak of the spectrum is always at smaller scales  
than standard CDM predictions, or observations.  
However the overall normalization of the spectrum increases 
considerably as $w^X$ decreases. 
 
The CDM rms fluctuation in 8 $h^{-1}$Mpc spheres is $\sigma_8=0.4, 
0.6, 1.8$ for $w^X=1/3,0.1,0.01$. Hence relativistic decay products 
match well the observed $\sigma_8\approx 0.5$. On the other hand 
in 100 $h^{-1}$Mpc spheres one requires bias $b_{100}= 
\sigma_{100}^{data}/\sigma_{100}=4.9, 3.7, 1.6$ 
to match observations.   
 
Energy dumps into radiation have no effect on the CDM power spectrum. 
However if there is energy transfer into CDM or baryons, even  
with $w^X=1/3$, the CDM power spectrum is highly enhanced. 
This is due to the addition of small scale entropy fluctuations 
to the usual fluctuations gravitationally induced by the stings. 
We plot the result of a 5\% transfer into CDM and a 20\%  
transfer into baryons (with $w^X=1/3$)   
for which $b_{100}=2.0, 1.5$. 
 
Hence in our calculations local strings have a bias problem 
at 100 $h^{-1}$Mpc, although its magnitude is not 
as great as found in \cite{abr}.  It depends sensitively on  
the decay products, being reduced if the strings have a channel  
into non-relativistic particles, 
or if there is some energy transfer into the baryon and CDM fluid. 
The main problem with strings in an $\Omega=1$, $\Omega_b=0.05$,  
$\Omega_\Lambda = 0$ CDM Universe  
is that the  
shape of $P(k)$ never seems to match observations. This may not 
be the case with other cosmological parameters \cite{abr2,avelino}.   
 
\section{Strings and inflation}

One way of improving upon the previous situation is to consider 
mixed scenarios: strings and inflation. 
Recent developments in inflation model building, based on supersymmetry,
have produced compelling models in which strings are 
produced at the end of inflation.
In such models the cosmological perturbations are 
seeded both by the defects and by the quantum fluctuations.

A major drawback of inflationary theories 
is that they are far-removed from particle physics
models. Attempts to improve on this state of affairs have been made
recently, resorting to supersymmetry 
\cite{Cas+89,DTermInfl,rachel,LytRio98,Cop+94,Tka+98,sug}.
In these models one identifies
flat directions in the potentials, which are enforced by a (super)symmetry. 
Such flat directions produce ``slow-roll inflation''. 
In order to stop inflation one must tilt the potential, allowing 
for the fields to roll down.  
In so-called D-term supersymmetric inflationary scenarios, 
inflation stops with a symmetry-breaking phase transition, 
at which a U(1) symmetry is spontaneously broken, leading to the 
formation of cosmic strings.  This is only the most 
natural of a whole class of models of so-called hybrid inflation.
Hence a network of cosmic strings is formed at the end of inflation.

\begin{figure}[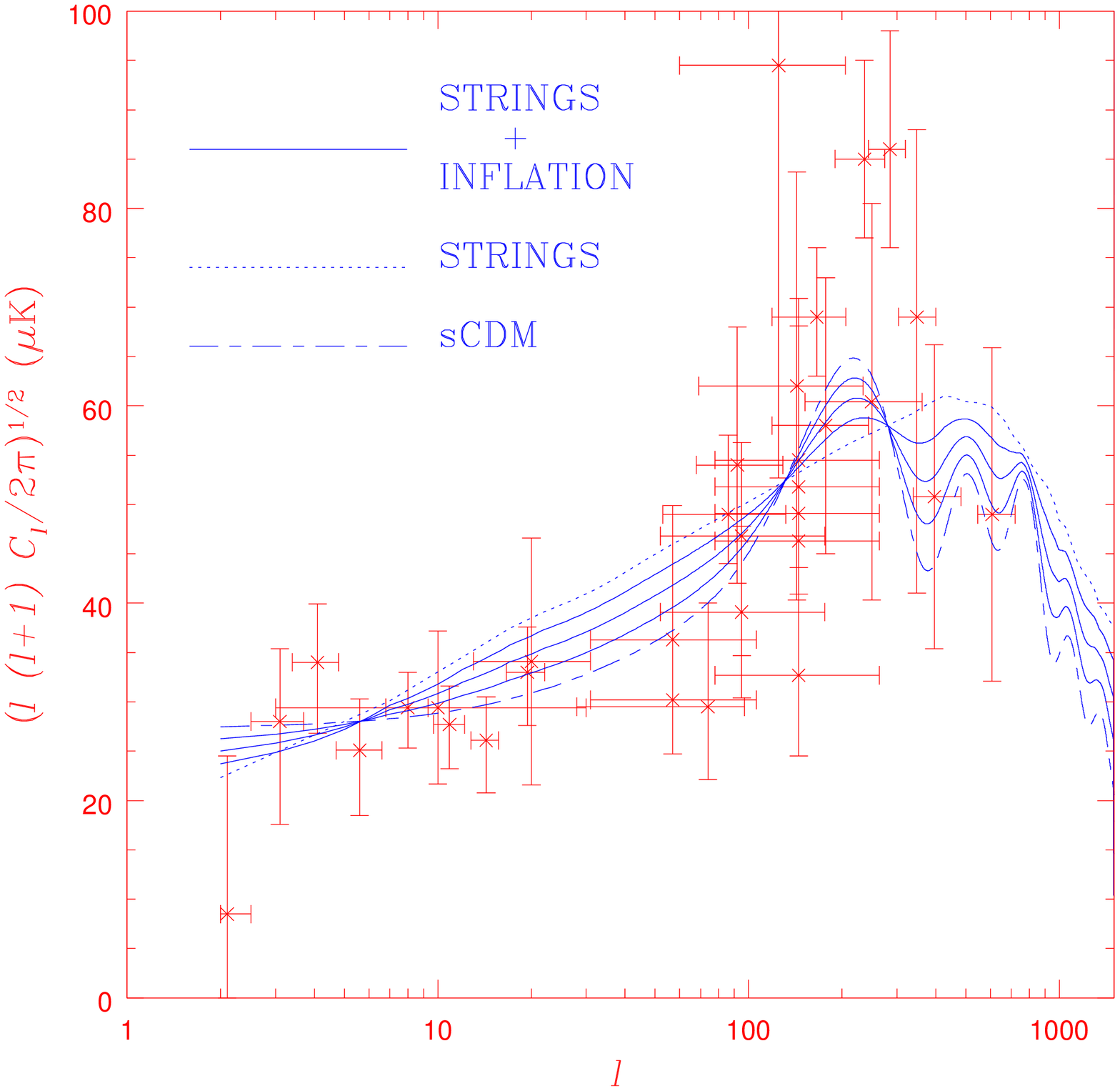]
\begin{center}
\leavevmode \epsfysize=8cm  \epsfbox{res3.ps}
\end{center}
\caption{\label{res3}The CMB power spectra predicted by cosmic strings, sCDM, 
and by inflation and strings with
$ R_{\rm{SI}}=0.25,0.5,0.75.$ 
The large angle
spectrum is always slightly tilted. The Doppler peak becomes a thick
Doppler bump at $\ell=200-600$, modulated by mild undulations.} 
\end{figure}

\begin{figure}[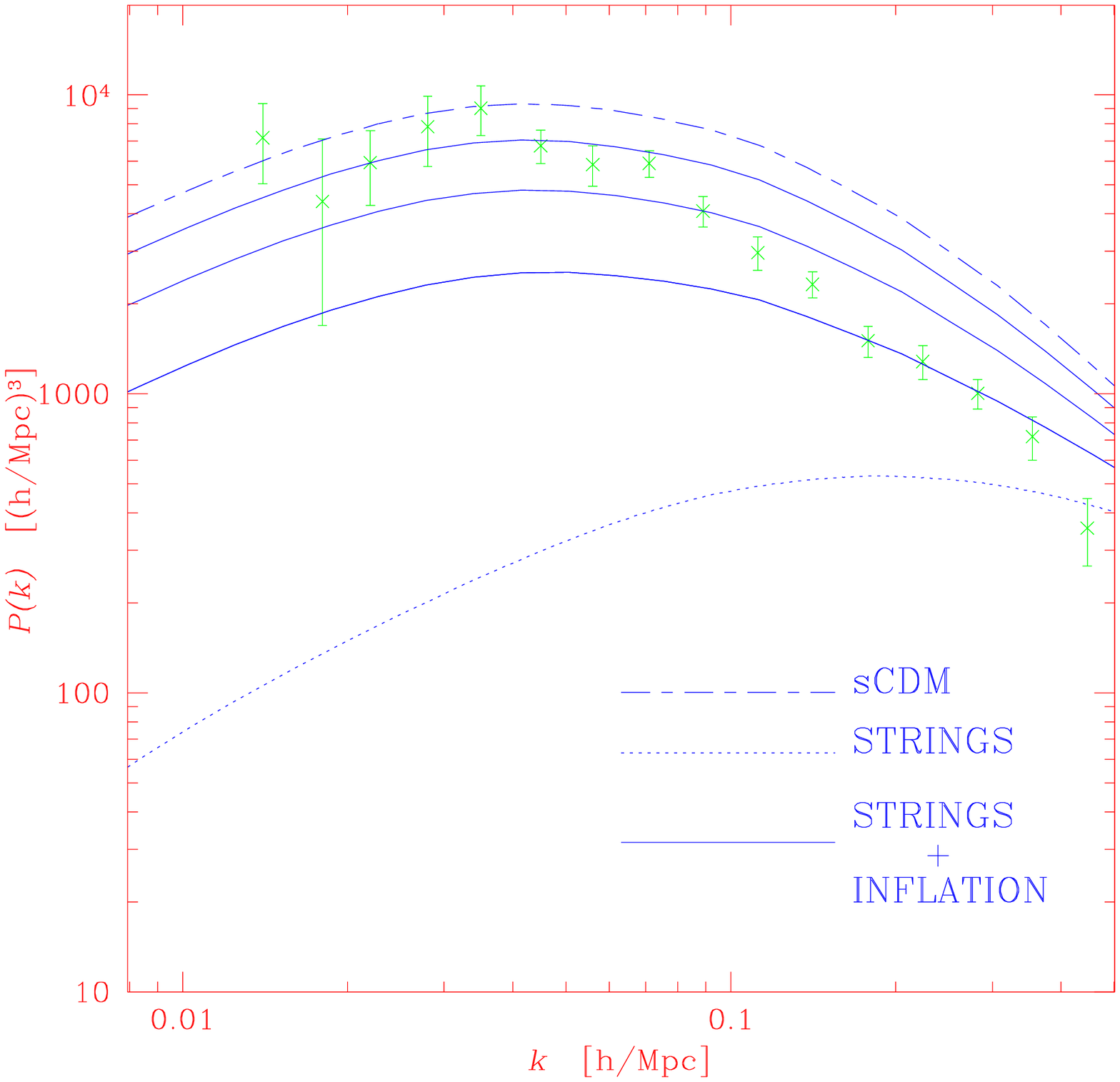]
\begin{center}
\leavevmode \epsfysize=8cm  \epsfbox{res4.ps}
\end{center} 
\caption{\label{res4} The  power spectra in CDM fluctuations 
predicted by cosmic strings, sCDM, 
and by inflation and strings with $R_{\rm{SI}}=0.25,0.5,0.75.$
We have also superposed the power spectrum as inferred from surveys
by Peacock and Dodds.} 
\end{figure}

In Figs. (\ref{res3}) and (\ref{res4})
 we present power spectra in CMB and CDM produced 
by a sCDM scenario, by cosmic strings, and by strings plus inflation. 
We have assumed the traditional choice of parameters, setting the Hubble 
parameter
$H_0=50$ km sec$^{-1}$ Mpc$^{-1}$, the baryon fraction to
$\Omega_b=0.05$, and  assumed a flat geometry, no cosmological
constant, 3 massless neutrinos, standard recombination,
and cold dark matter. The inflationary perturbations have a 
Harrison-Zel'dovich or scale invariant spectrum, and the amount of
gravitational radiation (tensor modes)
produced during inflation is assumed to be negligible.

We now summarise the results.
\begin{itemize}
\item 
The CMB power spectrum shape in these models
is highly exotic. The inflationary 
contribution is close to being Harrison-Zel'dovich. Hence it produces
a flat small $\ell$ CMB spectrum. The admixture of strings, 
however, imparts a tilt.
Depending on $R_{\mathrm{SI}}$ (the ratio of large angle anisotropy
due to strings and inflation)
one may tune the CMB plateau tilt between 1
and about 1.4, without invoking primordial tilt and inflation
produced gravity waves.

\item
The proverbial inflationary Doppler peaks are transfigured in these
scenarios into a thick Doppler bump, covering the region  
$\ell=200-600$. The height of the peak is similar for sCDM and strings,
with standard cosmological parameters. 
The Doppler bump is modulated by small undulations, 
which cannot truly be called secondary peaks. 
By tuning $R_{\mathrm{SI}}$ one may achieve any degree
of secondary oscillation softening. This provides a major
loophole in the argument linking inflation with secondary
oscillations in the CMB power spectrum \cite{andbar,inc}. 
If these oscillations were not
observed, inflation could still survive, in the form of the models
discussed above.

\item
In these scenarios the LSS of 
the Universe is almost all produced by inflationary fluctuations. 
However COBE scale CMB anisotropies are due to both strings and  
inflation. Therefore COBE normalized CDM fluctuations are  
reduced by a factor $(1+R_{\mathrm{SI}})$ in strings plus inflation scenarios. 
This is equivalent to multiplying 
the sCDM bias by ${\sqrt{1+R_{\mathrm{SI}}}}$ on all scales, except
the smallest, where the string contribution may be
non negligible. Given that 
sCDM scenarios produce too much structure on small scales
(too many clusters)  
this is a desirable feature. 
\end{itemize}

Overall ``Strings plus inflation'' are interesting first of all as an inflationary
model. Its ``flat potential'' is not the result of a finely tuned coupling
constant, but the result of a symmetry. Hence in some sense
these models achieve inflation without fine tuning. The only free parameters
are the number of inflationary e-foldings, and the scale of symmetry
breaking. These parameters also fix the absolute (and therefore relative)
normalizations of string and inflationary fluctuations.

The combination of these two scenarios smoothes 
the hard edges of either separate component, leaving 
a much better fit to LSS and CMB power spectra. We illustrated this
point in this review, but left
out a couple of issues currently under investigation
which we now summarise.

The CDM power spectrum in these scenarios has a break at very small
scales, when string produced CDM fluctuations become dominant over
inflationary ones. This aspect was particularly emphasized in
\cite{stinf2}, and there is some observational evidence in favour
of such a break. An immediate implication of this result is that
it is easier to form structure at high redshifts \cite{steidel,lalfa}. 
In \cite{bmw} it is shown that even with Hot Dark Matter, these
scenarios produce enough damped
Lyman-$\alpha$ systems, to account for the recent high-redshift
observations.

Another issue currently under investigation is the timing of structure
formation \cite{steidel}. Active models drive fluctuations
at all times, and therefore produce a time-dependence
in $P(k)$ different from passive models. The effect is subtle,
but works so as to slow down structure formation. Hence for
the same normalization nowadays there is more structure at high
redshifts in string scenarios \cite{AB95,MB96,ZLB}.

\section{Defects and Baryogenesis}

Baryogenesis forms another overlap area between defects in particle physics and cosmology. The goal is to explain the observed asymmetry between matter and antimatter in the Universe. In particular, the objective is to be able to deduce the observed value of the net baryon to entropy ratio at the present time
\begin{equation}
{{\Delta n_B} \over s}(t_0) \, \sim \, 10^{-10} 
\end{equation}
starting from initial conditions in the very early Universe when this ratio vanishes. Here, $\Delta n_B$ is the net baryon number density and $s$ the entropy density.

As pointed out by Sakharov \cite{Sakharov}, three basic criteria must be satisfied in order to have a chance at explaining the data:
\begin{enumerate}
\item{} The theory describing the microphysics must contain baryon number violating processes.
\item{} These processes must be C and CP violating.
\item{} The baryon number violating processes must occur out of thermal equilibrium.
\end{enumerate}

As was discovered in the 1970's \cite{GUTBG}, all three criteria can be satisfied in GUT theories. In these models, baryon number violating processes are mediated by superheavy Higgs and gauge particles. The baryon number violation is visible in the Lagrangian, and occurs in perturbation theory (and is therefore in principle easy to calculate). In addition to standard model CP violation, there are typically many new sources of CP violation in the GUT sector. The third Sakharov condition can also be realized: After the GUT symmetry-breaking phase transition, the superheavy particles may fall out of thermal equilibrium. The out-of-equilibrium decay of these particles can thus generate a nonvanishing baryon to entropy ratio. 

The magnitude of the predicted $n_B / s$ depends on the asymmetry $\varepsilon$ per decay, on the coupling constant $\lambda$ of the $n_B$ violating processes, and on the ratio $n_X / s$ of the number density $n_X$ of superheavy Higgs and gauge particles to the number density of photons, evaluated at the time $t_d$ when the baryon number violating processes fall out of thermal equilibrium, and assuming
that this time occurs after the phase transition. The quantity $\varepsilon$ is proportional to the CP-violation parameter in the model. In a GUT theory, this CP violation parameter can be large (order 1), whereas in the standard electroweak theory it is given by the CP violating phases in the CKM mass matrix and is very small. As shown in \cite{GUTBG} it is easily possible to construct models which give the right $n_B / s$ ratio after the GUT phase transition (for recent reviews of baryogenesis see \cite{Dolgov} and \cite{RubShap}).
 
The ratio $n_B / s$, however, does not only depend on $\varepsilon$, but also on $n_X / s (t_d)$. If the temperature $T_d$ at the time $t_d$ is greater than the mass $m_X$ of the superheavy particles, then it follows from the thermal history in standard cosmology that $n_X \sim s$. However, if $T_d < m_X$, then the number density of $X$ particles is diluted exponentially in the time interval between when $T = m_X$ and when $T = T_d$. Thus, the predicted baryon to entropy ratio is also exponentially suppressed:
\begin{equation} \label{expdecay}
{n_B \over s} \, \sim \, {1 \over {g^*}} \lambda^2 \varepsilon e^{- m_X / T_d} \, ,
\end{equation}
where $g^*$ is the number of spin degrees of freedom in thermal equilibrium at the time of the phase transition.
In this case, the standard GUT baryogenesis mechanism is ineffective.

However, topological defects may come to the rescue \cite{BDH}. As was discussed at the beginning of these lecture notes, topological defects will inevitably be produced in the symmetry breaking GUT transition provided they are topologically allowed in that symmetry breaking scheme. The topological defects provide an alternative mechanism of GUT baryogenesis in the following way:
Inside of topological defects, the GUT symmetry is restored. In fact, the defects can be viewed as solitonic configurations of $X$ particles. The continuous decay of defects at times after $t_d$ provides an alternative way to generate a nonvanishing baryon to entropy ratio. The defects constitute out of equilibrium configurations, and hence their decay can produce a nonvanishing $n_B / s$ in the same way as the decay of free $X$ quanta. 

The way to estimate the $n_B / s$ ratio is as follows: The defect
scaling solution gives the energy density in defects at all times. Taking the time derivative of this density, and taking into account the expansion of the Universe, we obtain the loss of energy attributed to defect decay. By energetics, we can estimate the number of decays of individual quanta which the defect decay corresponds to. We can then use the usual perturbative results to compute the resulting net baryon number.

Provided that $m_X < T_d$, then at the time when the baryon number violating processes fall out of equilibrium (when we start generating a nonvanishing $n_B$) the energy density in free $X$ quanta is much larger than the defect density, and hence the defect-driven baryogenesis mechanism is subdominant. However, if $m_X > T_d$, then as indicated in (\ref{expdecay}), the energy density in free quanta decays exponentially. In contrast, the density in defects only
decreases as a power of time, and hence can soon dominate baryogenesis.

One of the most important ingredients in the calculation is the time dependence of $\xi(t)$, the separation between defects. Immediately after the phase transition at time $t_f$ (when the defect network is formed), the separation is $\xi(t_f) \sim \lambda^{-1} \eta^{-1}$. In the time period immediately following, the time period of relevance for baryogenesis, $\xi(t)$ approaches the Hubble radius according to the equation \cite{Kibble2} 
\begin{equation} \label{defsep}
\xi(t) \, \simeq \, \xi(t_f) ({t \over {t_f}})^{5/4} \, .
\end{equation}
Using this result to calculate the defect density, we obtain after some algebra
\begin{equation} \label{barres}
{{n_B} \over s}|_{\rm defect} \, \sim \, \lambda^2 {{T_d} \over \eta} {{n_B} \over s}|_0 \, ,
\end{equation}
where $n_B / s|_0$ is the unsuppressed value of $n_B / s$ which can be obtained using the standard GUT baryogenesis mechanism. We see from (\ref{barres}) that even for low values of $T_d$, the magnitude of $n_B / s$ which is obtained via the defect mechanism is only suppressed by a power of $T_d$. However, the maximum strength of the defect channel is smaller than the maximum strength of the usual mechanism by a geometrical suppression factor $\lambda^2$ which expresses the fact that even at the time of defect formation, the defect network only occupies a small fraction of the volume of space.

It has been known for some time that there are baryon number violating processes even in the standard electroweak theory. These processes are, however, non-perturbative. They are connected with the t'Hooft anomaly \cite{tHooft}, which in turn is due to the fact that the gauge theory vacuum is degenerate, and that the different degenerate vacuum states have different quantum numbers (Chern-Simons numbers). In theories with fermions, this implies different baryon number. Configurations such as sphalerons \cite{sphal} which interpolate between two such vacuum states thus correspond to baryon number violating processes.

As pointed out in \cite{KRS85}, the anomalous baryon number violating processes are in thermal equilibrium above the electroweak symmetry breaking scale. Therefore, any net baryon to entropy ratio generated at a higher scale will be erased, unless this ratio is protected by an additional quantum number such as a nonvanishing $B - L$ which is conserved by electroweak processes.

However, as first suggested in \cite{Shap} and discussed in detail in many recent papers (see \cite{EWBGrev} for reviews of the literature), it is possible to regenerate a nonvanishing $n_B / s$ below the electroweak symmetry breaking scale. Since there are $n_B$ violating processes and both C and CP violation in the standard model, Sakharov's conditions are satisfied provided that one can realize an out-of-equilibrium state after the phase transition. Standard model CP violation is extremely weak. Thus, it appears necessary to add some sector with extra CP violation to the standard model in order to obtain an appreciable $n_B / s$ ratio. A simple possibility which has been invoked often is to add a second Higgs doublet to the theory, with CP violating relative phases. 

The standard way to obtain out-of-equilibrium baryon number violating processes immediately after the electroweak phase transition is \cite{EWBGrev}  to assume that the transition is strongly first order and proceeds by the nucleation of bubbles (note that these are two assumptions).

Bubbles are out-of-equilibrium configurations. Outside of the bubble (in the false vacuum), the baryon number violating processes are unsuppressed, inside they are exponentially suppressed. In the bubble wall, the Higgs fields have a nontrivial profile, and hence (in models with additional CP violation in the Higgs sector) there is enhanced CP violation in the bubble wall. In order to obtain net baryon production, one may either use fermion scattering off bubble walls \cite{CKN1}  (because of the CP violation in the scattering, this generates a lepton asymmetry outside the bubble which converts via sphalerons to a baryon asymmetry) or sphaleron processes in the bubble wall itself \cite{TZ,CKN2}. It has been shown that, using optimistic parameters (in particular a large CP violating phase $\Delta \theta_{CP}$ in the Higgs sector) it is possible to generate the observed $n_B / s$ ratio. The resulting baryon to entropy ratio is of the order
\begin{equation} \label{ewres}
{{n_B} \over s} \, \sim \, \alpha_W^2 (g^*)^{-1} \bigl( {{m_t} \over T} \bigr)^2 \Delta \theta_{CP} \, ,
\end{equation}
where $\alpha_W$ refers to the electroweak interaction strength, $g^*$ is the number of spin degrees of freedom in thermal equilibrium at the time of the phase transition, and $m_t$ is the top quark mass. The dependence on the top quark mass enters because net baryogenesis only appears at the one-loop level.

However, analytical and numerical studies show that, for the large Higgs masses which are indicated by the current experimental bounds, the electroweak phase transition will unlikely be sufficiently strongly first order to proceed by bubble nucleation. In addition, there are some concerns as to whether it will proceed by bubble nucleation at all (see e.g. \cite{Gleiser}).

Once again, topological defects come to the rescue. In models which admit defects, such defects will inevitably be produced in a phase transition independent of its order. Moving topological defects can play the same
role in baryogenesis as nucleating bubbles. In the defect core, the electroweak symmetry is unbroken and hence sphaleron processes are unsuppressed \cite{Perkins}, provided that the core is sufficiently thick to contain the sphalerons (in recent work \cite{Cline} it has been shown that this is a rather severe constraint on workable string-mediated electroweak baryogenesis mechanisms). In the defect walls there is enhanced CP violation for the same reason as in bubble walls. Hence, at a given point in space, a nonvanishing baryon number will be produced when a topological defect passes by.

Defect-mediated electroweak baryogenesis has been worked out in detail in \cite{BDPT} (see \cite{BD} for previous work) in the case of cosmic strings. The scenario is as follows: at a particular point $x$ in space, antibaryons are produced when the front side of the defect passes by. While $x$ is in the defect core, partial equilibration of $n_B$ takes place via sphaleron processes. As the back side of the defect passes by, the same number of baryons are produced as the number of antibaryons when the front side of the defect passes by. Thus, at the end a positive number of baryons are left behind. Obviously, the advantage of the defect-mediated baryongenesis scenario is that it does not depend on the order and on the detailed dynamics of the electroweak phase transition.

As in the case of defect-mediated GUT baryogenesis, the strength of defect-mediated electroweak baryogenesis is suppressed by the ratio ${\rm SF}$ of the volume which is passed by defects divided by the total volume, i.e.
\begin{equation}
{{n_B} \over s} \, \sim \, {\rm SF} {{n_B} \over s}|_0 \, ,
\end{equation}
where $(n_B / s)|_0$ is the result of (\ref{ewres}) obtained in the bubble nucleation mechanism. 

A big caveat for defect-mediated electroweak baryogenesis is that the standard electroweak theory does not admit topological defects. However, in a theory with additional physics just above the electroweak scale it is possible to obtain defects (see e.g. \cite{TDB95} for some specific models). The closer the scale $\eta$ of the new physics is to the electroweak scale $\eta_{EW}$, the larger the volume in defects and the more efficient defect-mediated electroweak baryogenesis (however, as pointed out in \cite{Cline}, this effect is counteracted by the fact that the defect velocity at $T = \eta_{EW}$ decreases as $\eta$ decreases). Using the result of (\ref{defsep}) for the separation of defects, we obtain (for non-superconducting strings)
\begin{equation}
{\rm SF} \, \sim \, \lambda \bigl( {{\eta_{EW}} \over \eta} \bigr)^{3/2} v_D\, .
\end{equation} 
where $v_D$ is the mean defect velocity. Typically \cite{Cline}, the resulting value of $SF$ is too small for string-mediated electroweak baryogenesis to be efficient.
 
Defect-mediated baryogenesis is much more efficient if the defects are domain walls (since by purely geometrical arguments the factor $SF$ will be much larger). However, as seen in earlier sections of this review, theories with topologically stable domain walls are ruled out because the wall network would overclose the Universe. What one would like is a theory in which a network of walls forms at some time before the electroweak phase transition, remains present until after the transition and then decays. As has recently been
realized \cite{NB99}, this goal may be achieved in theories with embedded defects.

Embedded defects (see \cite{AV99} for a recent review) in a theory with vacuum manifold ${\cal M}$ are exact solutions of the field equations which correspond to topological defects with respect to a submanifold ${\cal M}'$ of ${\cal M}$
of strictly lower dimension. Embedded defects are unstable in the absence of field fluctuations. However \cite{NB99} if plasma effects provide effective masses to some of the components of the order parameter of the theory such that the space of ground states in the plasma reduces to ${\cal M}'$, then the embedded defects can be stabilized in the plasma. Embedded strings which can be stabilized in the electric plasma before recombination exist in the standard electroweak theory. It is interesting to investigate if extension of the minimal electroweak theory admit embedded walls. Such walls would then yield ideal
candidates for defect-mediated electroweak baryogenesis. 

It is important to keep in mind that defects occurring at any scale may have
important consequences for baryogenesis which should be considered when
exploring the cosmological implications of the models which admit such defects.

\end{document}